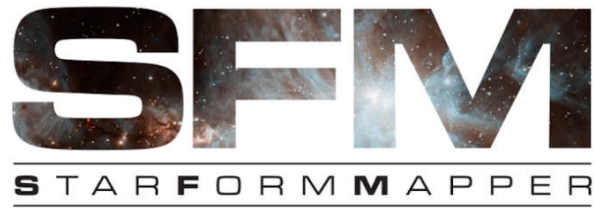

# D1.1
# Report on Optimal Substructure Techniques for Stellar, Gas and Combined Samples

| | |
|---|---|
| Prepared by: | I. Joncour, A. Buckner, P. Khalaj, E. Moraux, F. Motte |
| Lead Beneficiary: | UNIVERSITE JOSEPH FOURIER GRENOBLE 1 |
| Version Number: | 1 |
| Date of Issue: | (06/06/2017) |
| Status:[1] | Complete |
| Distribution:[2] | PU |
| Document Type:[3] | R |
| Approved by: | Stuart Lumsden |

[1] *Draft, Approved*
[2] *PU = Public; CO = Confidential, only for members of the Consortium and EU.*
[3] *R = Report; R+O = Report plus Other. Note: all "O" deliverables must be accompanied by a deliverable report.*

Document History

| Version Number | Date | Editor | Modification |
|---|---|---|---|
| 1 | (06/06/2017) | Anne Buckner | Updated, PDF created. |
| | | | |
| | | | |

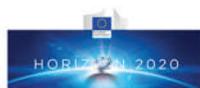

*This project is being funded by the European Union's Horizon 2020 research and innovation actions (RIA) programme under the grant agreement No 676036.*

# Contents

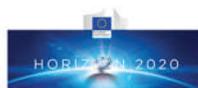

*This project is being funded by the European Union's Horizon 2020 research and innovation actions (RIA) programme under the grant agreement No 676036.*

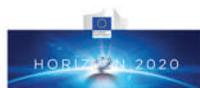

*This project is being funded by the European Union's Horizon 2020 research and innovation actions (RIA) programme under the grant agreement No 676036.*

# List of Figures

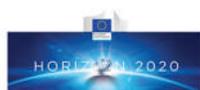

*This project is being funded by the European Union's Horizon 2020 research and innovation actions (RIA) programme under the grant agreement No 676036.*

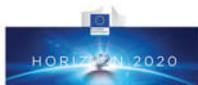

*This project is being funded by the European Union's Horizon 2020 research and innovation actions (RIA) programme under the grant agreement No 676036.*

# 1. General Introduction

## 1.1. Scope

This document is intended to review the different types of clustering algorithm in order to identify the best type to study the spatial and kinematic clustering of stars. It is the deliverable "D1.1 Report on Optimal Substructure Techniques for Stellar, Gas and Combined Samples" for the EU H2020 (COMPET-5-2015 – Space) project "A Gaia and Herschel Study of the Density Distribution and Evolution of Young Massive Star Clusters" (Grant Agreement Number: 687528), with abbreviated code name StarFormMapper (SFM) project. The structure and content is based on [AD1].

The document is organised in the following sections:

1. General Introduction
2. Clustering of Discrete Distributions
3. Clustering of Continuous Distributions
4. Clustering in Astrophysics
5. StarFormMapper
6. Summary & Conclusions

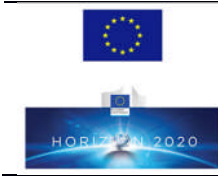

*This project is being funded by the European Union's Horizon 2020 research and innovation actions (RIA) programme under the grant agreement No 676036.*

## 1.2. Acronyms

| | |
|---|---|
| 2dFGRS | 2dF Galaxy Redshift Survey |
| AIC | Akaike Information Criterion |
| BIC | Bayesian Information Criteria |
| CDM | Cold Dark Matter |
| CMP | Curvature Mask Pixels |
| CUPID | ClUmP IDentification and analysis package |
| CUTEX | CUrvature Thresholding EXtractor |
| DBSCAN | Density-Based Spatial Clustering of Applications with Noise |
| DENCLUE | DENsity CLUstering |
| DD | Documents Deliverable |
| EM | Expectation-Maximisation |
| eMST | Euclidean MST |
| ESA | European Space Agency |
| FOF | Friends-Of-Friends |
| FWHM | Full Width at Half Maximum |
| GB | GigaByte |
| GDBSCAN | Generalized DBSCAN |
| GDL | GNU Data Language |
| IDL | Interactive Data Language |
| KDD | Knowledge Database Discovery |
| MB | MegaByte |
| MST | Minimum Spanning Tree |
| MYStIX | Massive Young Star-formIng compleX |

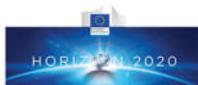

*This project is being funded by the European Union's Horizon 2020 research and innovation actions (RIA) programme under the grant agreement No 676036.*

| NDF | extensible N-dimensional Data Format |
|-----|--------------------------------------|
| NESTS | Nested Elementary STructureS |

(Acronyms continued from previous page)

| OPTICS | Ordering points to identify the clustering structure |
|--------|------------------------------------------------------|
| PSF | Point Spread Function |
| rms | root-mean-square |
| RV | Radial Velocity |
| SDSS | Sloan Digital Sky Survey |
| SFM | StarFormMapper |
| SKG | Spectrum of Kinematic Grouping |
| SExtractor | Source Extractor |
| TGAS | Tycho-Gaia Astrometric Solution |
| YSO | Young Stellar Object |

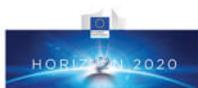

*This project is being funded by the European Union's Horizon 2020 research and innovation actions (RIA) programme under the grant agreement No 676036.*

# 2. Clustering of Discrete Distributions

Clustering is the process of examining a collection of objects embedded in a p-dimensional space to automatically group the closest objects together into natural clusters whilst separating distinct groups from each other. Although this task can (sometimes) be easily done by eye in a 2D space, it is not straightforward to formalise the process on general grounds. The primary problems with formalisation are: (1) defining, conceptually and quantitatively, what constitutes 'close' in a given distribution and hence to define from heterogeneous variables a proximity/similarity or a distance/dissimilarity function between objects; (2) defining a meaningful cluster membership criteria to (or not) assign objects in clusters.

Consequently a plethora of clustering algorithms have been developed since the sixties to identify clustering in discrete distributions, guided by and tailored to, suit the data and objectives of author's studies (for a general reviews on clustering methods, see e.g. Everitt et al., 2011; Jain, 2010; Bhattacharyya and Hazarika, 2007; Roy and Bhattacharyya, 2005; Tan et al., 2005).

However, these algorithms can be broadly placed into two category types: *Hierarchical* methods (agglomerative and divisive, minimum spanning tree) and *Partitioning* methods (including density-based, grid-based and model-based methods). In this Section we provide a comprehensive overview of these two basic types of algorithms.

## 2.1. Hierarchical Clustering

Standard hierarchical clustering groups data over a variety of scales by creating a *cluster binary tree* or *dendrogram*. The tree is not a single set of clusters, but rather a multilevel hierarchy, where clusters at one level are joined as clusters at the next level. If the user is free to state the scale of clustering that appears to be the 'most appropriate' (i.e. decide the cut-off level), they are required to decide and quantitatively define 'appropriate'. Of course, this definition is a matter of debate since there is no general law, so at best the criteria is heuristic, and ultimately will depend on the clustering purpose of the user.

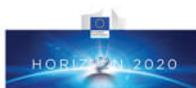

*This project is being funded by the European Union's Horizon 2020 research and innovation actions (RIA) programme under the grant agreement No 676036.*

### 2.1.1. Agglomerative Algorithms

A traditional class of hierarchical clustering algorithms are the so-called *agglomerative* algorithms. These non-parametric algorithms start from initial set of clusters (each composed of a single object) and iteratively build nested families of larger clusters through successive merges of cluster pairs, using a heuristic proximity function criteria, until a single cluster containing all objects in the distribution is obtained.

The other traditional class of clustering algorithms are the non-parametric *divisive* hierarchical clustering algorithms (Kaufman and Roussee 1990, which essentially are the reverse of the agglomerative algorithms i.e. starting from the whole set and step by step dividing the set in subsets until single objects are obtained (but the techniques are a bit more complicated, see Filippone et al., 2008 for a review on spectral partitioning). However as these are much less used, and as the same ideas apply to both divisive and agglomerative algorithms, this report will focus on the agglomerative algorithms. (For a full review on hierarchical clustering, the reader is directed to Everitt et al. (2011)).

In agglomerative algorithms a proximity function is defined heuristically at each merging step as described by the *linkage method* being employed, which is selected by the user prior to implementing the algorithm. Arguably, the six most widely employed linkage methods (as illustrated in Figure 1) are:

◆ Single Link The single link method estimates the distance D between two clusters as that of the closest pair of all pairs consisting of one member of each cluster *(D= min{d(x,y) : x ∈A, y ∈B}).*

◆ Complete Link The complete link method estimates the distance D between two clusters as that of the furthest pair of all pairs consisting of one member of each cluster *(D= max{d(x,y) : x ∈A, y ∈B}).*

◆ Average Link The average link method estimates the distance D between two clusters as the mean of all the distances between the pairs consisting of one member of each cluster *(D= mean{d(x,y) : x ∈A, y ∈B}).*

◆ Centroid Link The centroid link method estimates the distance D between two clusters as that of the centroids (means) of each cluster *(D= {d($x_C$,$y_C$) : $x_C$ barycentre of A, $y_C$ barycentre of B}).*

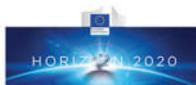

*This project is being funded by the European Union's Horizon 2020 research and innovation actions (RIA) programme under the grant agreement No 676036.*

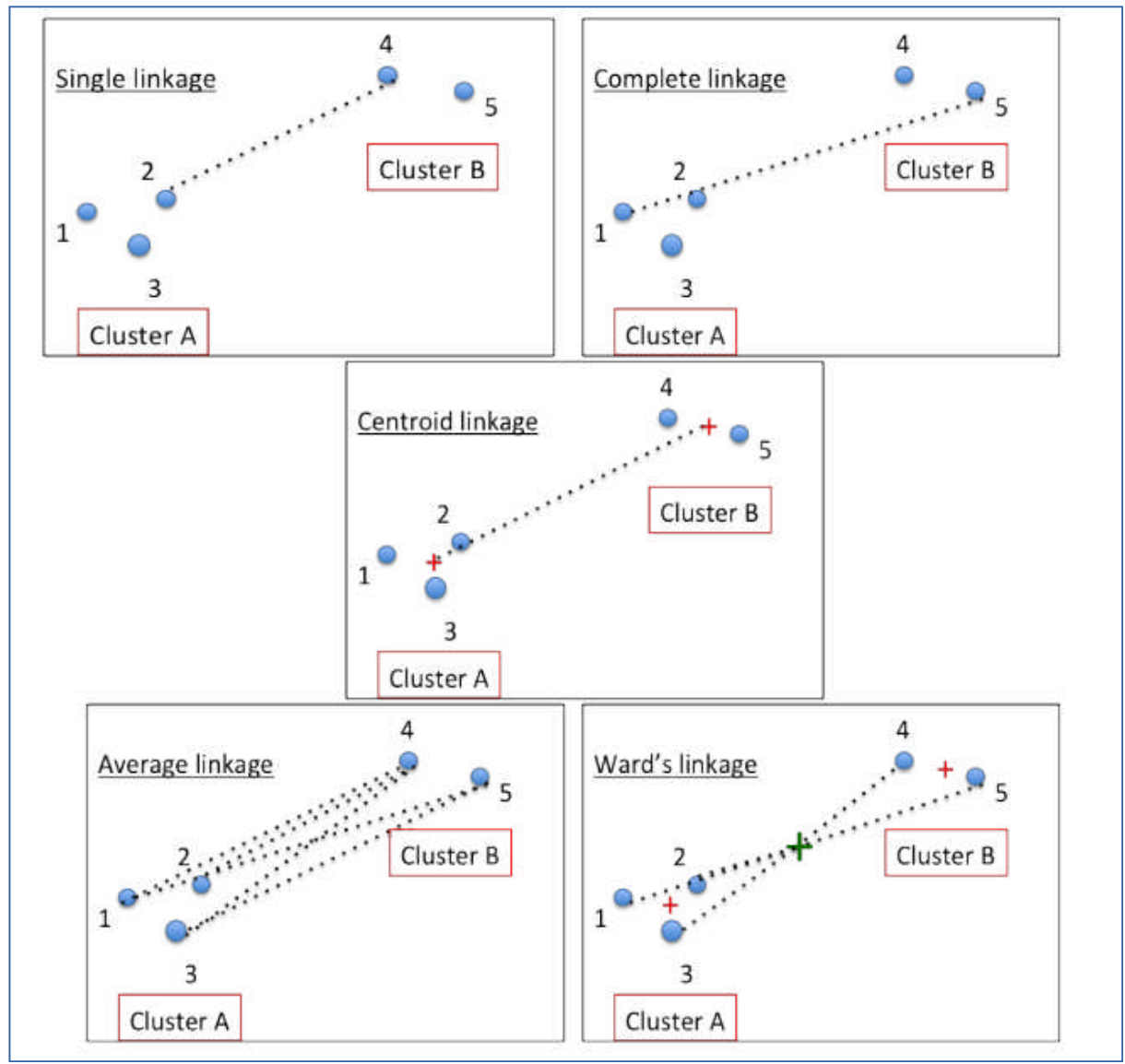

Figure 1: Illustration of the six most widely employed linkage methods in agglomerative hierarchical clustering (see text for details).

This project is being funded by the European Union's Horizon 2020 research and innovation actions (RIA) programme under the grant agreement No 676036.

◆ Median Link As the centroid link is unfair in the case of the merging of two unbalanced clusters (merging of one large cluster with a small one), instead of taking the centroids of the cluster as the reference points, the mid-point of two centres of the clusters obtained during the previous merging is chosen.

◆ Ward's Method This method estimates the distance D between two clusters as the sum of squared distances from each points to the centroid of two clusters in the set $(D = \mathrm{sum}\{d^2(z, \mu_c) : z \in A \cup B, \mu_c \text{ barycentre of } A \cup B\})$.

In each method the cluster pair in the set with the smallest distance function, D, (named also dissimilarity function) are then merged. The similarity of two elements within a cluster is quantified with a distance measure, and different type of metrics can be used to determine such a distance. It can be either the usual Euclidean distance or other distances such as Manhattan, Mahalanobis, or Hamming.

The strength of these non-parametric methods is that no assumptions are made about the structure of data, such as the a priori number of clusters that has to be given in partitioning methods (Section 2.2). However, they suffer from a significant drawback: since each step of agglomerative process is built upon the previous one, and no backtracking permitted, there is a loss of data such that the nested hierarchical structure is built-in and not the result of an open process free of reconfiguring the structure of data at each step. In other words, the hierarchical clustering algorithm is sensitive to possibly erroneous previous cluster merging, since (once assigned) object's cluster memberships are not permitted to change i.e. assignments at each step are permanent.

Another significant drawback are the linkage methods lack of robustness, especially with respect to noise and outliers in distributions, which presents as either spurious additional clusters being identified and/or the blurring of distinct clusters (Figure 2). It also tends to produce clusters of a particular shape. For example, in the single linkage method the merging of two clusters relies only on a *local* merging criteria based on their *most proximate* members. Thus only the closest parts of the two clusters are considered whereas the overall structure of the clusters and their more distant parts are not taken into account. Since clusters are merged on the criteria of the closest element pair of two clusters in a set, it may happen that in a noisy distribution elements are successively merged in such a way that two obvious distinct clusters are merged together (due to few sparse singletons being

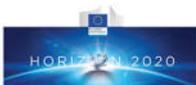

*This project is being funded by the European Union's Horizon 2020 research and innovation actions (RIA) programme under the grant agreement No 676036.*

close to each other), even though many of the elements in each cluster are very distant from each other. This tendency to combine elements linked by a series of close intermediate elements gives rise to long chains (this is known as the *chaining effect*).

On the other hand in the complete linkage methods the merging of two clusters is based on the proximity of their furthest members. Therefore at each step a merged cluster that has the smallest diameter is created, and since the diameter of the merged clusters is minimised at each iteration, the method has a tendency to give rise to compact "globular" clusters. This criterion for merging in complete link methods is non-local in the sense that the entire distribution of members in a cluster can influence the choice of the merging. For example, a single element located at a relatively large distance from the majority of the other members of a cluster will significantly increases the final diameter of two merged clusters, which in turn may lead to a major change in the final clustering. Thus the complete-link methods are rather sensitive to outliers, as the single-link methods are sensitive to the noise.

## 2.1.2. Minimum Spanning Trees

Hierarchical clustering groups data over a variety of scales by creating a cluster tree or dendrogram. The tree is not a single set of clusters, but rather a multilevel hierarchy where clusters at one level are joined as clusters at the next level. Closely related to the hierarchical methods (especially the single link algorithm) the Minimum Spanning Tree (MST) method is a graph theoretic procedure (Kruskal, 1956; Prim, 1957; Dijkstra, 1960) to derive clusters from a data point set (Zahn, 1971).

When considering a set of (resp. spatial) points as a complete graph, in which the points are vertices and edges are the (resp. Euclidean) distance between vertices, the MST (resp. eMST, i.e. Euclidean MST) is the subgraph in which any two vertices are connected by exactly one simple path (i.e. a connected graph without cycle, by definition a tree) such that the total distance length of its edges is minimal. The method generates clusters by deleting the largest lengths of MST above a heuristically chosen threshold (inconsistent edges: Zahn, 1971, or the critical length: Gutermuth et al., 2009), such that clusters are defined as the remaining connected subgraphs (i.e. there is a path between any two points of the subgraph).

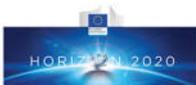

*This project is being funded by the European Union's Horizon 2020 research and innovation actions (RIA) programme under the grant agreement No 676036.*

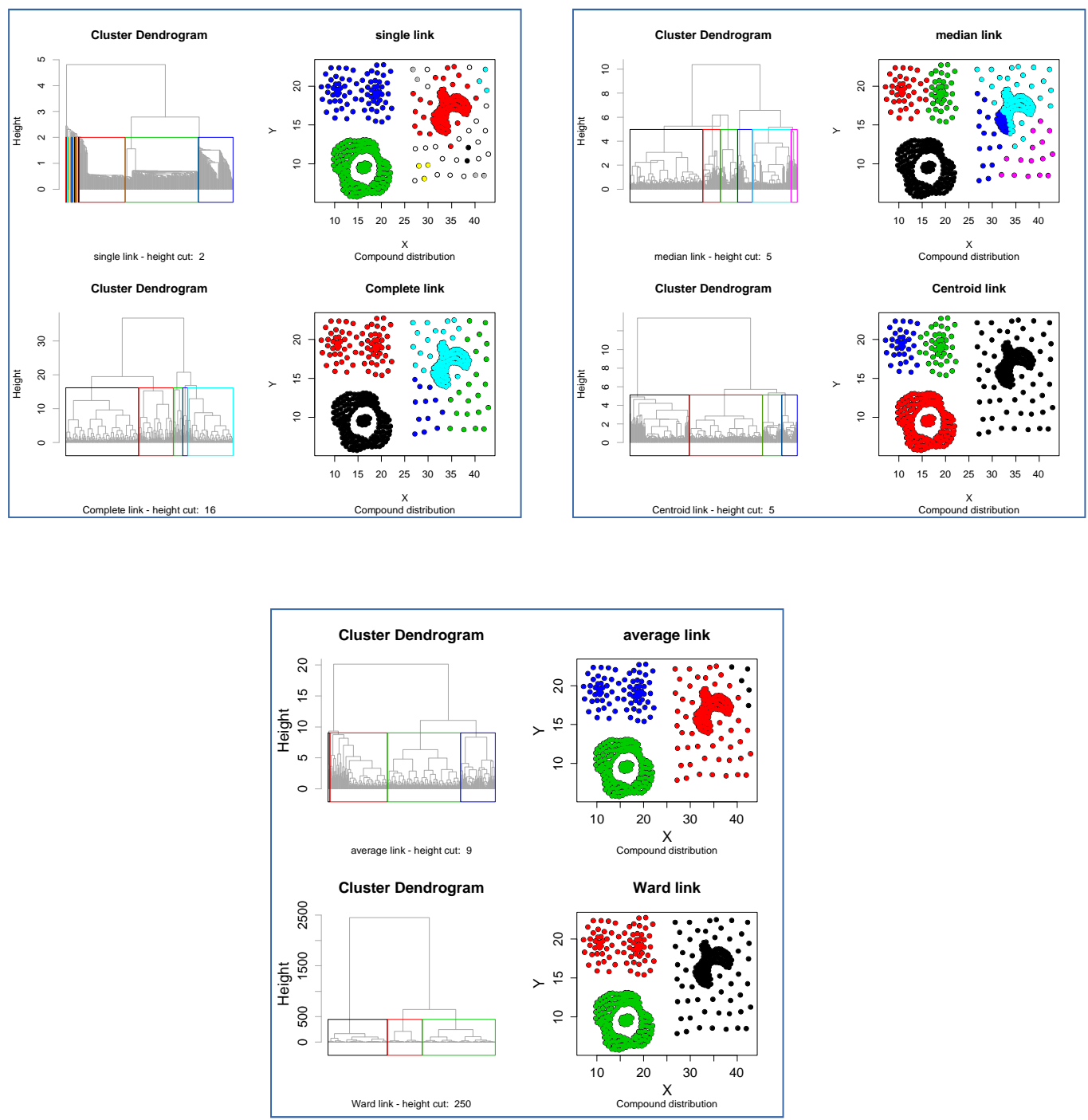

Figure 2  Demonstrates how the number of clusters identified and the dendrogram generated for a given same data set varies with the linkage method employed, shown here for the single, complete, median, centroid, average and Ward link methods.

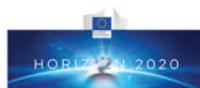

*This project is being funded by the European Union's Horizon 2020 research and innovation actions (RIA) programme under the grant agreement No 676036.*

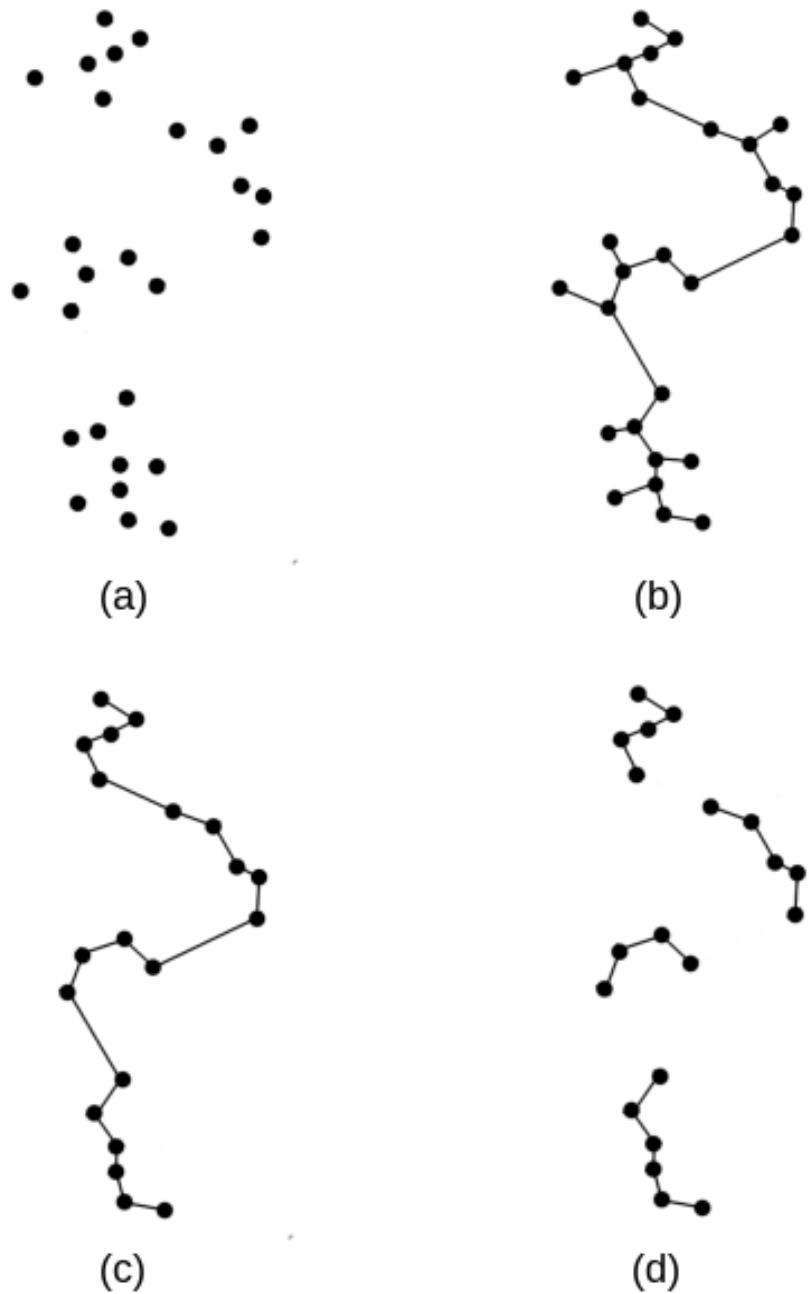

Figure 3 Shows the construction and reduction of a minimal spanning tree (MST). (a) A point data set, (b) MST constructed according to the prescription given in Section 2.1.2 of the text, (c) pruned MST; all nodes in (b) of degree one connected to nodes with degree exceeding two have been removed along with their connecting edges, (d) Separate and pruned MST; all edges of (c) exceeding a critical length have been removed.

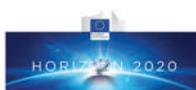

*This project is being funded by the European Union's Horizon 2020 research and innovation actions (RIA) programme under the grant agreement No 676036.*

## 2.2. Partition Clustering

Partition clustering algorithms group a discrete distribution into clusters of traditionally non overlapping subsets by building partitions in the dataset into a set of clusters. These partitions are constructed using either (1) iterative relocation, (2) density or (3) grid based criteria.

### 2.2.1. Iterative Relocation Based

Iterative relocation algorithms use an iterative control strategy to optimize the partition of distribution into clusters. The number of clusters is usually an input parameter for these algorithms, i.e. some a priori domain assumption knowledge is required (for a review, see Chapter 3 in Mirkin (2005))

Once the number of clusters is given, there are two statistical approaches in order to partition the data in these clusters:

◆ Non-parametric Non-parametric iterative relation algorithms such as K-means or K-medoid, are based on the assumption that clusters correspond to different modes of the probability density, i.e. the values that appear more often in a data set. Once an initial partition is given, each cluster is associated to its representative e.g. its barycentre, centroid, mean point, or by one of the objects of the cluster located near its centre. The goal of this class of algorithm is to assign each object to its closest representative and iteratively process to minimize the objective function, i.e. the distance of elements to the their closest representatives summed over the over whole set; the iterative process stops when convergence on representatives is reached.

For example, the K-means algorithm (Hartigan, 1975; Hartigan and Wong, 1979; Macqueen, 1967), as a variance-based function, can be shown to minimize within-cluster distance while maximizing between-cluster distance. It should be noted that the assignment of each object to a cluster implies that the induced partition is

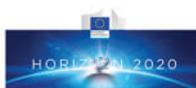

*This project is being funded by the European Union's Horizon 2020 research and innovation actions (RIA) programme under the grant agreement No 676036.*

equivalent to a Voronoi diagram. Thus the shape of clusters found by a partitioning algorithm is always convex, which is a restrictive bias of the method. A second drawback is that there is no advice for choosing the number of clusters and locations of the initial seeds. The final clusters are usually close to the initial seeds so that the results much depend on initialization.

◆   Parametric Parametric iterative relation algorithms (mixture models) are based on the assumption that each cluster is represented by a set of parametric probability density coming from a same family. For example, a Gaussian distribution would be employed for a Gaussian mixture model method. This type of clustering is also called model-based clustering. They have been developed to deal with the structure identification based on the optimization of the Bayesian Information Criteria (BIC) that allows the choice of the best model based on direct models comparison, models that are computed simultaneously, each combining different weight of clustering methods (hierarchical and EM partitioning) and various number of components (Fraley and Raftery, 1998, 2002, 2007). However, even these models do not answer to the question of the possible skewness of data distribution that may lead to several normal distributions to describe one single skewed or elongated cluster. Moreover, if data is substructured, and since there is no explicit notion of noise, the best model will be chosen based on complete partition of data, which may be a serious obstacle in determining locally small but significant overdensities.

For this class of iterative relocation (resp. parametric) algorithm, the goal is to estimate: (1) the number (resp. and type of the geometrical distribution) of the clusters (structure identification); and (2) the parameters of the distributions (weight, variance and mean for each probability distribution associated to a cluster). Partitioning cluster analysis algorithms are then based on two distinct phases: firstly, a model fitting phase, whereby the number and the geometrical type of component are a priori chosen; and secondly, a model validation phase, whereby the set of data are assessed to each cluster according to some cluster validity criterion and to one optimal partition hypothesis selected through a maximisation technique of one objective function. Most of the time the Expectation-Maximisation (EM) technique of the (log)-likelihood function is used. The EM technique is able to propose the best local solution in order to softly distribute data in each cluster and to constrain the parameters of each probability distribution associated with a cluster. For example, the K-means approach is a special case of EM clustering applied to a mixture of

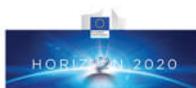

*This project is being funded by the European Union's Horizon 2020 research and innovation actions (RIA) programme under the grant agreement No 676036.*

Gaussians when all variances are equal (Gupta and Chen, 2011), i.e. it is another way of seeing why the K-means method leads to globular shaped clusters. 

Iterative relocation methods have several drawbacks: the number of clusters or components have to be a priori to implementation, they are sensitive to first initialization partition, the derived clusters are biased toward convex/ globular shaped clusters, they don't identify arbitrarily sized clusters (biased to equipartition), and the notion of noisy data is not considered i.e. it is a termination requirement that every single object has to be assigned to a cluster. In summary, the structure iterative relocation algorithms impose on, and their required complete partitioning of, discrete distributions limit the identification of non-globular, irregularly shaped and local small-scale substructures.

## 2.2.2. Density Based

Density clustering algorithms partition a distribution into clusters based on a density criterion. Essentially they are the formalisation of the assumption that clusters are high-density regions surrounded by low-density regions. These algorithms borrow (1) a part of the non-parametric and parametric iterative partition cluster methods, and (2) a part of MST (single linkage) hierarchical clustering methods.

For the latter, clusters are also defined as the maximal connected subgraphs remaining at a cutting scale length threshold, but these subgraphs are composed by selected instances (thus allowing an incomplete partitioning).

For the former, density clustering algorithms are a non-parametric method, as there is no priori given function associated to cluster postulating a specific structure for discrete distributions. However the notion of density probability is not completely ruled out since there is an estimation of the local density from a kernel density estimator associated to a given smoothing bandwidth that reduces the complexity. The non-parametric estimation of probability density functions is based on the concept that the value of a density function at a continuity point can be estimated using the sample observations that fall within a small region around that point, known as the Parzen kernel class of density estimates. (For a comprehensive review of kernel non-parametric density estimation and the crucial issue of the smoothing parameter (bandwidth) choice see e.g. Scott, 1992; Silverman, 1986).

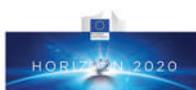

*This project is being funded by the European Union's Horizon 2020 research and innovation actions (RIA) programme under the grant agreement No 676036.*

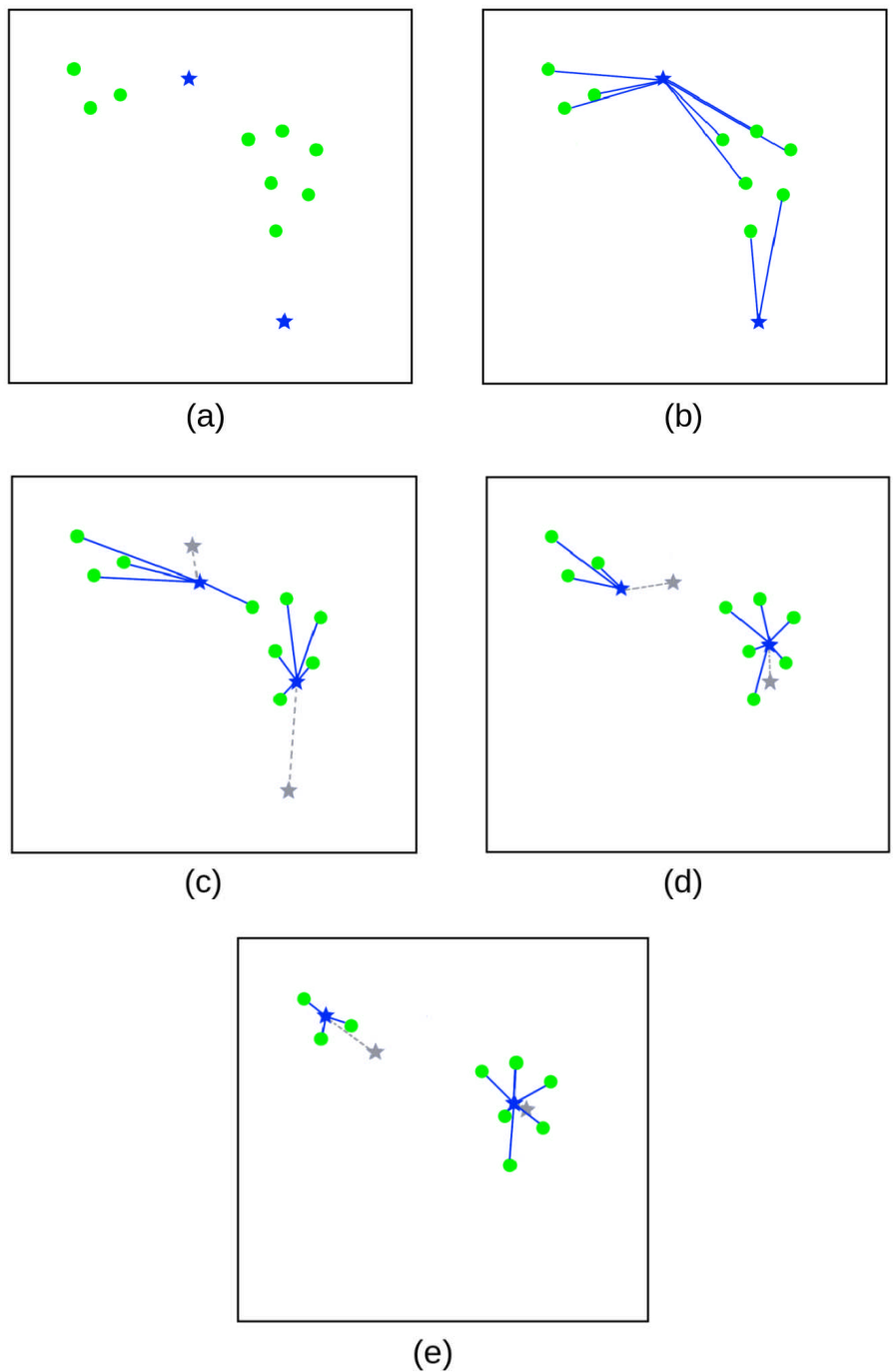

Figure 4 Illustrated example of the Iterative Relocation Based Partition algorithm "K-means". The first step starts with an initial guess of the centroids (a), then the data are grouped to their nearest centroids (b). In the next step (c) the new centroids are determined as the centres of these new clusters and the points are regrouped to the nearest (new) centroid. This process is iteratively repeated [(d)] until the clusters remain the same [(e)].

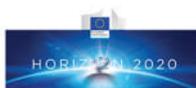
*This project is being funded by the European Union's Horizon 2020 research and innovation actions (RIA) programme under the grant agreement No 676036.*

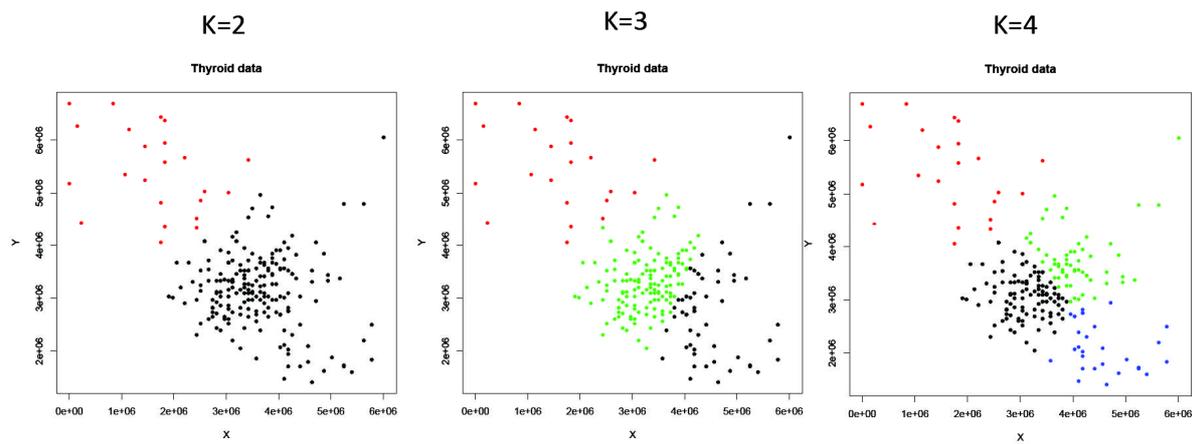

Figure 5 Demonstrates how the identified clustering of a data set with the K-means algorithm is dependent on the number of centroids, K, specified by the user, shown here for (left) K=2; (middle) K=3; and (right) K=4. Colours represent the membership of points to the identified clusterings.

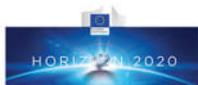

*This project is being funded by the European Union's Horizon 2020 research and innovation actions (RIA) programme under the grant agreement No 676036.*

Below we discuss the three most widely used density based algorithms.

◆ DBSCAN (Density-based spatial clustering of applications with noise)

This algorithm is a powerful tool to partition the data in clusters based on the connectivity of the points having a similar neighbourhood density (Ester et al.,1996). Given a set of points in space, it groups together points that: (1) are close neighbours (separated by a distance less than ε) and (2) have in common the same minimum local density, i.e. that have at least the same minimum of points $N_{min}$ within a sphere of radius ε around each point. If this last condition is fulfilled for the two points, the points are said to be core-points and directly (density-)reachable.

DBSCAN is one of the most common clustering algorithms and also the most cited in scientific literature. Two points in a cluster are said to be simply (density-)reachable if there is a path between these 2 points where each point along the path is directly reachable from the previous point. All the points along the path have then to be core-points, except the two extreme points at the ends of the path for which the condition on the minimal local density may not be fulfilled; they are then called border points. A point that is not reachable from any other point is called an outlier. A cluster is formed by all the points that are reachable from all the other points (core or border points), but contain at least one core point (see Figure 6).

With respect to the other partitioning algorithms previously described, DBSCAN does not require one to specify the number of clusters in the data a priori. DBSCAN can find arbitrarily shaped clusters. Due to the $N_{min}$, condition, it allows to filter the noise in the data, and the so-called chaining effect as seen in single-link clustering (different clusters being connected by a thin line of points) is reduced. The distribution of points in clusters do not depend on the ordering of points, except possibly for the border points. Since a border point is assigned to a cluster at the first occurrence. There are two free parameters that can be set depending on the goal that the user intends to reach. The often cited main drawback is its poor ability to deal with clusters of different densities, but this can be handled by setting the local free parameters (ε, $N_{min}$) to values that are suitable to the detection of the less dense cluster in such a way that, since they are local upper thresholds, they are also suitable to detect denser clusters.

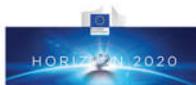

*This project is being funded by the European Union's Horizon 2020 research and innovation actions (RIA) programme under the grant agreement No 676036.*

A popular extension of DBSCAN is GDBSCAN (Generalized DBSCAN) which generalizes the notion of point density and therefore it can be applied to objects of arbitrary data type, e.g. 2-dimensional polygons (Ankerst et al., 1999).

◆ OPTICS (Ordering Points To Identify the Clustering Structure)

While DBSCAN computes a single level clustering, i.e. clusters of single user defined density, the OPTICS algorithm represents the intrinsic, hierarchical structure of the data by a (one-dimensional) ordering of the points (Sander et al., 1998). The resulting graph (called reachability plot; see Figure 7), which is a representation in between the silhouette as defined by Rousseeuw (1987) and the dendrograms as obtained by the hierarchical algorithms previously discussed, visualizes clusters of different densities as well as hierarchical clusters. It was then further developed as F-OPTICS (F-OPTICS) to introduce fuzzy clustering in the case of uncertain data (Kriegel and Pfeifle, 2005).

OPTICS is capable to detect irregularly shaped variable density clusters; however, it fails to detect nested or embedded clusters. Moreover, OPTICS requires a prior ordering of objects in terms of reachability distance, which incurs additional cost. To address this problem, EnDBSCAN was proposed in Roy and Bhattacharyya (2005) which is an enhanced version of DBSCAN and OPTICS. EnDBSCAN can detect any embedded or intrinsic clusters efficiently. It extends the concept of core distance used in OPTICS and introduced the concept of core neighbourhood which enables to handle the problem of global density parameter setting, suffered by DBSCAN. Furthermore, like other well-known density based approaches, it also gives the number of clusters naturally, in presence of noise.

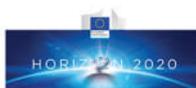

*This project is being funded by the European Union's Horizon 2020 research and innovation actions (RIA) programme under the grant agreement No 676036.*

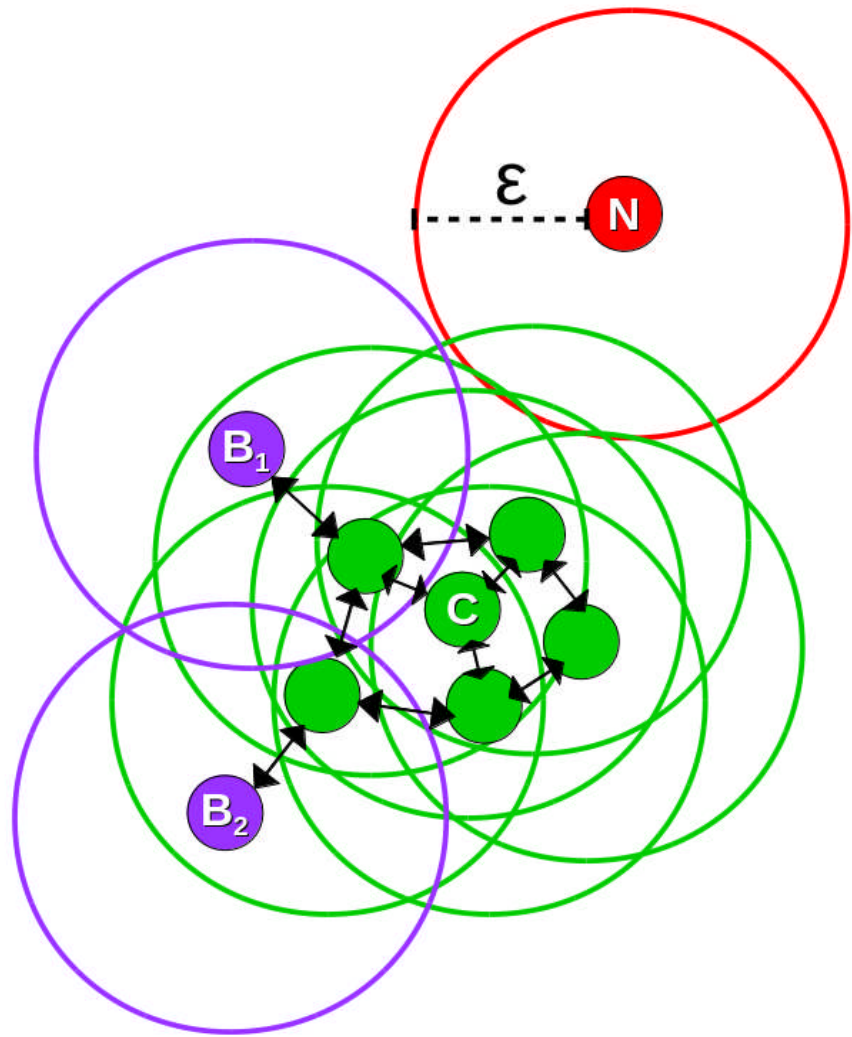

Figure 6 Illustrative example of the DBSCAN algorithm applied to a data set with 9 points and $N_{min}$=4. Points are colour coded as follows: (green) core points; (purple) border points; (red) noise points. A core points of the cluster (such as C) **contains a minimum 4 points (including the point itself) within a surrounding radius of ε.** $B_1$ and $B_2$ are density reachable from C via core points so are border points of the cluster, whilst point N is not density reachable from core points so is designed as Noise.

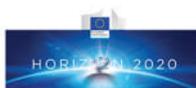

*This project is being funded by the European Union's Horizon 2020 research and innovation actions (RIA) programme under the grant agreement No 676036.*

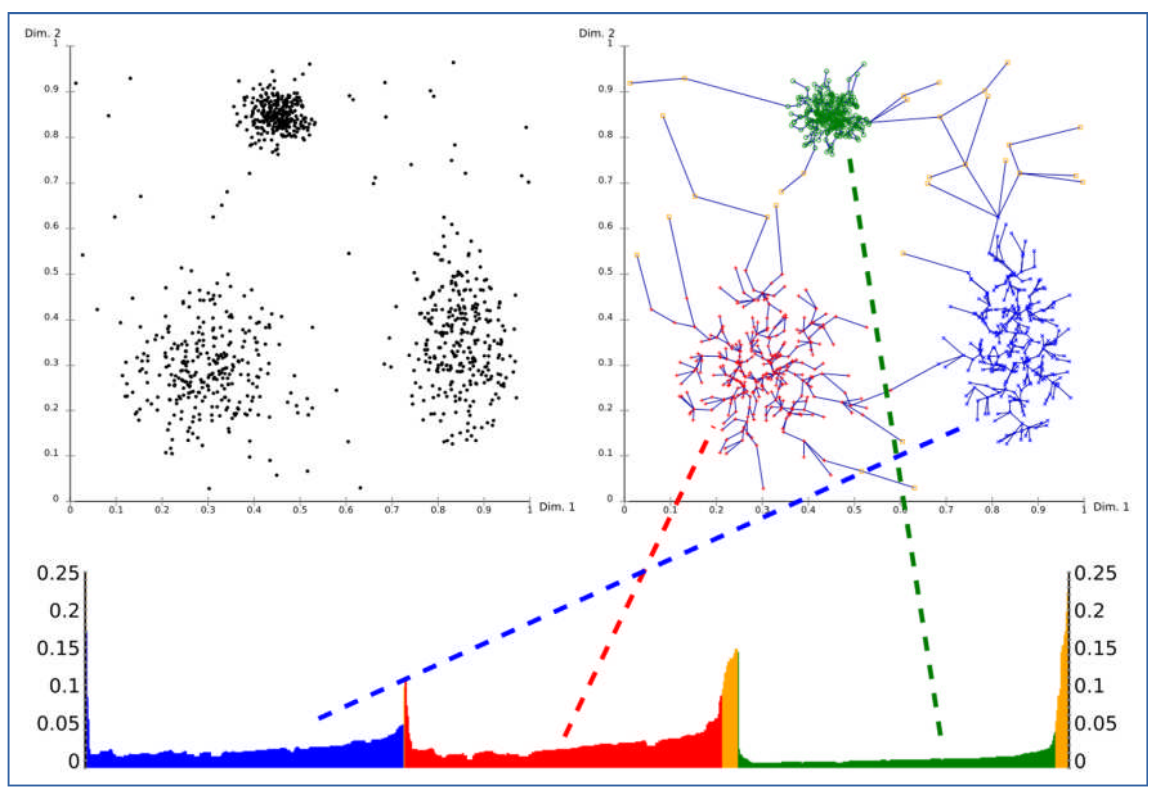

**Figure 7 Example of the OPTICS algorithm applied to a data set with parameters ε>0.5 and Nmin=10. (Top Left:)** Data set before algorithm is applied: (Top Right:) clusters identified by OPTICS and cluster order (connecting each point with its predecessor): (Bottom:) the reachability plot showing order of points [x-axis] against reachability distance [y-axis][1].

---

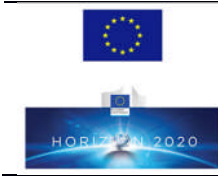

*This project is being funded by the European Union's Horizon 2020 research and innovation actions (RIA) programme under the grant agreement No 676036.*

◆ DENCLUE (DENsity CLUstering)

The DENCLUE algorithm (Hinneburg and Gabriel, 2007; Hinneburg and Keim, 1998) is a density based partitioning clustering technique that uses a parametric cluster model based on kernel density estimation. A cluster is defined by a local maximum of the estimated density function. Data points are assigned to clusters by hill climbing, i.e. points going to the same local maximum are put into the same cluster. It uses hill climbing procedure for Gaussian kernels (see Fig. 3 in Hinneburg and Gabriel, 2007), which adjusts the step size automatically at no extra costs. The procedure converges exactly towards a local maximum by reducing it to a special case of the expectation maximization algorithm.

A significant advantage of DENCLUE over DBSCAN is that it is significantly faster (unto a factor of 45; Hinneburg and Keim, 1998). However, as with the majority of clustering algorithms DENCLUE requires careful selection of its two input parameters: $\sigma$ (the influence of a point in its neighbourhood) and $\varepsilon$ (the minimum density level for a density-attractor to be significant) as their values can significantly affect the clustering found (see Fig. 4 in Hinneburg and Keim, 1998).

## 2.2.3. Grid Based

Grid clustering groups data by overlaying a grid structure over the data space to partition it into a finite number of cells. Closely related to the density based partition algorithm, the grid based methods are an alternate set of algorithms designed to build clusters from a given cell density threshold where a cluster is defined as a maximal set of neighbouring dense cells. Once the density of all cells has been computed clusters are created from the transversal of neighbouring dense cells and their central coordinates are calculated by identifying the densest cell(s) in the cluster.

This approach uses a multi-resolution grid data structure. For this purpose it quantizes the data space into a finite number of cells, forming the grid structure. The main advantage of the approach is its fast processing time, which depends only on the number of cells in each dimension of the quantized space, and not on the number of data objects. Approaches as STING, WaveCluster and CLIQUE are various examples of this approach and can be found in Han et al. (2011).

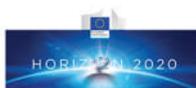

*This project is being funded by the European Union's Horizon 2020 research and innovation actions (RIA) programme under the grant agreement No 676036.*

The primary strengths of the grid based methods are that they are efficient (time complexity is typically O(n), for a data set of n objects) and scalable. Perhaps their greatest advantage is the ability to not only find clustering in *discrete* data sets but also *continuous* data sets (such as images) which can be achieved by using a wavelet transform to decomposes a signal into different frequency sub-bands, such that clusters are identified as connected components in the sub-bands data at different resolutions and scales.

Grid based approaches can have significant drawbacks: if a uniform grid structure, with predefined cell sizes, borders and cell density thresholds, is employed the algorithms can have significant difficulty in accurately finding clusters in highly irregular non-uniform data sets. However, the versatility of the grid-based approach is its adaptability to overcome these challenges. For example, high density data in a non-uniform distributions can be clustered if varying (adaptive) grid sizes are used in conjunction with axis-shifting partition strategy.

A number of adaptive grid based approaches are also closely related to the divisive hierarchical based algorithms (e.g. STING). This approach identifies clustering by storing summary statistics of grid cells analysed at different resolutions (called layers) in a hierarchical tree. For the first layer the data set is partitioned into a finite number of cells, for which their statistical parameters (number of objects; minimum, maximum, mean, and standard deviation of each dimension; type of distribution) are computed to find the cells of most relevance and stored in the hierarchical tree. For the second layer, each relevant first-layer cell is partitioned into a finite number of cells, for which their statistical parameters are computed to again find the cells of most relevance and are stored in the hierarchical tree. This recursive process is continued until the user-defined lowest layer is reached (see Figure 8).

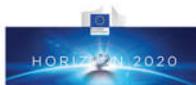

*This project is being funded by the European Union's Horizon 2020 research and innovation actions (RIA) programme under the grant agreement No 676036.*

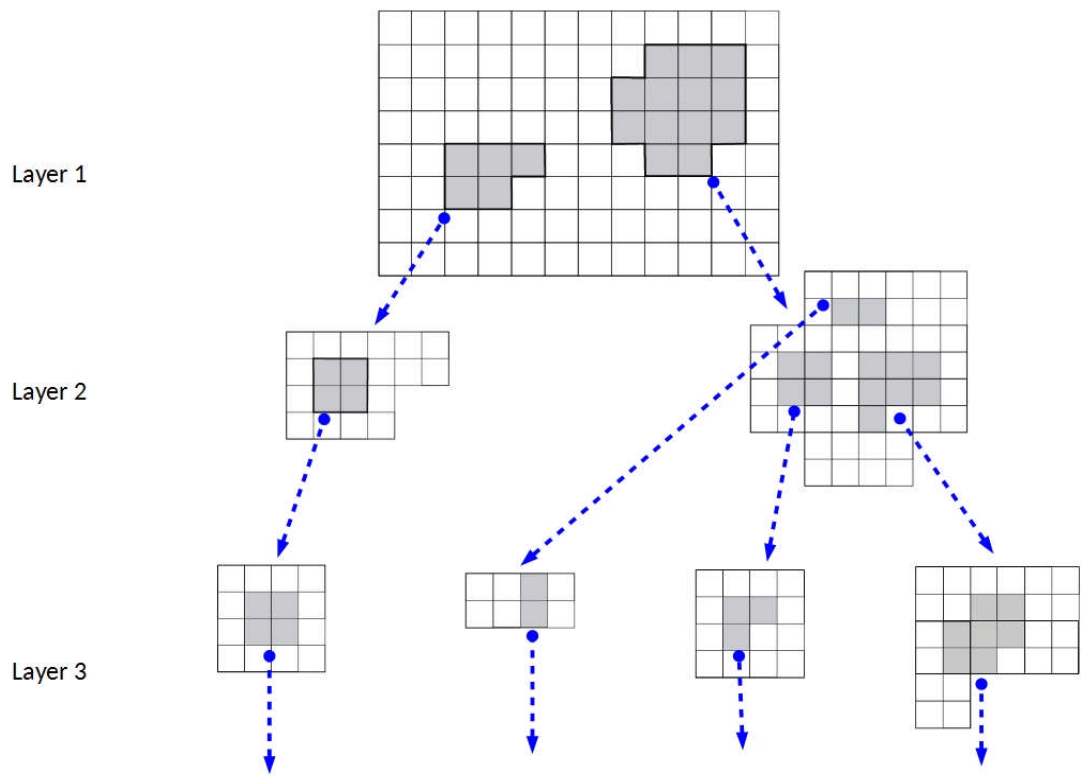

Layer 1

Layer 2

Layer 3

Figure 8 Illustration of adaptive grid clustering, the first layer is partitioned into a finite number of cells to find those of most relevance (grey cells). In the second layer these cells are further partitioned to find the most relevant ones. This recursive process is continued until the lowest layer is reached.

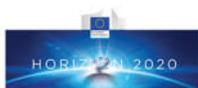

*This project is being funded by the European Union's Horizon 2020 research and innovation actions (RIA) programme under the grant agreement No 676036.*

## 2.3.  Assessment of Cluster Tendency and Validity

In order to verify whether the set of clusters obtained are reliable and correspond to a real phenomenon the validity of the clustering has to be checked. Are the clusters found in the data set 'true' or misidentified noise? Have the optimal (correct) number of clusters for the data set been found? Has the optimal clustering analysis/algorithm been used on the dataset?  For the latter question, different algorithms will produce different clustering results due to the methods they employ and (their discussed) associated biases in shape/type of cluster found (e.g. Figure 9).

For the first question, the primary assumption of any clustering analysis is that there is, in fact, clustering in the dataset and this can be dangerous i.e. clusters will be found in non-clustered (random) datasets. It is therefore prudent to asses the clustering tendency of the dataset (i.e. if the data set contains 'true' clusters) prior to running an analysis using the Hopkins statistic (Hopkins and Skellam (1954), Banerjee, A. (2004)). The statistic tests the spatial randomness of the data set by calculated and comparing the mean k-nearest neighbour distances between objects in randomly selected samples to that of a generated uniform data set, such that:

$$H = \frac{\sum_{i=1}^{n} v_i}{\sum_{i=1}^{n} u_i + \sum_{i=1}^{n} v_i}$$

Where H is the Hopkins statistic; n is the total number of objects in the randomly selected sample from data set X and the size of the generated uniform distribution Y; $v_i$ is the distance of $y_i \in Y$ from its nearest neighbour in X; $u_i$ is the distance of  $x_i \in X$ from its nearest neighbour in X. The value of H indicates whether the distribution of a data set is uniform (H=0), random (H=0.5) or highly clustered (H=1.0). A value of H>0.5 indicates clustered data and for H>0.75 a data set has a clustering tendency at the 90% confidence level (Lawson and Jurs, 1990).

Alternatively to (or in conjunction with) the Hopkins statistic a visual assessment, such as the Visual Assessment of cluster Tendency algorithm (VAT; Bezdek and Hathaway (2002)), can be used to asses clustering tendency.

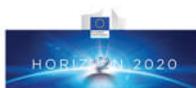

*This project is being funded by the European Union's Horizon 2020 research and innovation actions (RIA) programme under the grant agreement No 676036.*

Once the clustering tendency of the data set has been established and the chosen clustering analysis performed, the validity of the found clusters can be assessed using one (or more) of the following types of set criteria (called 'indexes') and the optimal number of clusters identified:

- ◆ Internal Index Internal indices asses the validity of clusters by comparing the similarity of assigned members to each other, and their dissimilarity to members of other clusters. This is done using a distance measure i.e. the compactness of a cluster, separation and homogeneity between clusters.

- ◆ External Index Unlike internal indices, external indexes asses the validity of clusters by comparing how well the found clusters match with an externally known result (e.g. cluster and class labels). The advantage of this method is that is it quantitatively quite easy to discern the validity of any given cluster (and specifically, the validity of different clustering), but on the other hand it requires additional information (variables) about a data set and therefore cannot be used on datasets where these external variables are unknown.

- ◆ Relative Index Relative indices asses the validity of clusters by varying the values of algorithms parameters to find the optimal number of clusters. This is particularly important for iterative based relocation algorithms such as k-means where the number of clusters is a user defined input value.

The choice of index type employed will depend on the dataset e.g. if no external information is known about the dataset a Internal and/or Random index will need to be used. Furthermore there is a plethora of implementations of the three types of indexes and their results may vary. It is therefore prudent when assessing the validity of clustering in a dataset to use more than one implementation. For a review and comparison of different validity index implementations the reader is referred to Milligan and Cooper (1985), Maulik and Bandyopadhyay (2002), Handl, Knowles, and Kell (2005), Rendón et al. (2011), Charrad et al., (2014), Li et al. (2016).

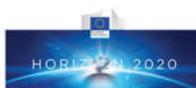

*This project is being funded by the European Union's Horizon 2020 research and innovation actions (RIA) programme under the grant agreement No 676036.*

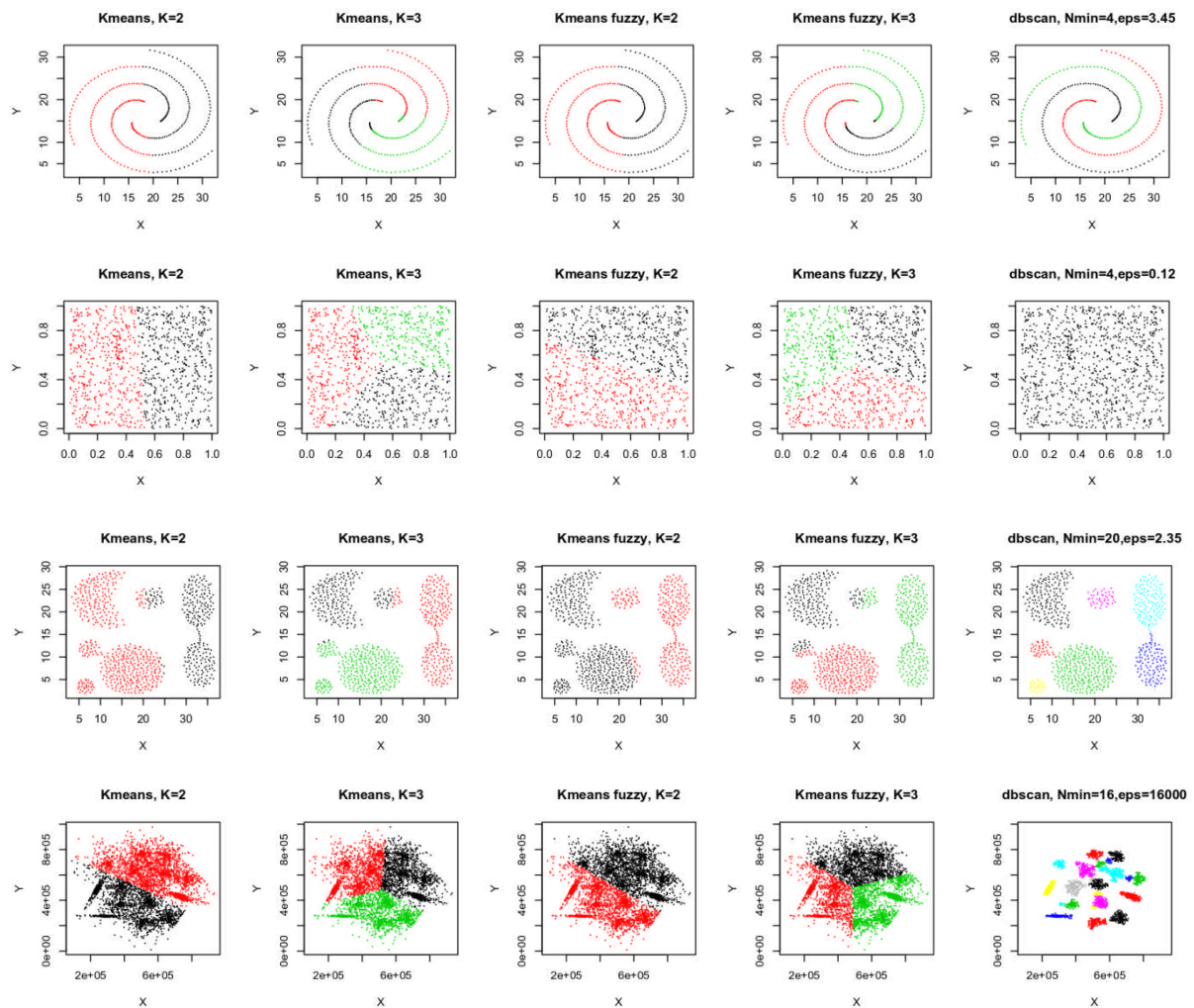

Figure 9 Illustration of clustering's identified using Iterative Relocation Based and Density Based algorithms on different data sets[2].

---

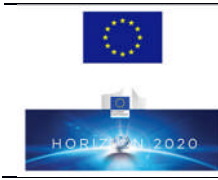

*This project is being funded by the European Union's Horizon 2020 research and innovation actions (RIA) programme under the grant agreement No 676036.*

# 3. Clustering of Continuous Distributions

Over the years several algorithms and computer programs have been developed to detect features and structures (e.g. sources, filaments and clumps) in continuous distributions such as molecular and gas clouds: CLUMPFIND (Williams, de Geus, and Blitz 1994); FELLWALKER (Berry 2015); GAUSSCLUMPS (Stutzki and Guesten 1990); CUPID (Berry et al. 2007, 2013; Wilson et al. 2011); REINHOLD (a part of the CUPID package); GETSOURCES (Men'shchikov et al. 2012a); CUTEX (Molinari et al. 2011); GETFILAMENTS (Men'shchikov 2013a); SExtractor (Bertin and Arnouts 1996a); tree structures and dendrograms (Houlahan and Scalo 1992; Rosolowsky et al. 2008a).

The aforementioned algorithms and computer programs are further elaborated upon below.

## 3.1. Algorithms

In the context of this report, there are three categories of algorithms used for detecting features in continuous distributions: clump identification, filament identification and studying hierarchical structures.

### 3.1.1. Clump Identification

#### 3.1.1.1. CLUMPFIND

This algorithm, as its name suggests, aims to identify clumps and is applicable to 2D as well as 3D data. The core of the algorithm relies on splitting the data into evenly spaced contours, e.g. at the multiples of the root-mean-square (rms) noise.

The algorithm starts by a contour level close to the value of the brightest peak (i.e. the pixel with the maximum value) and goes to smaller contour levels until it reaches the minimum contour level (i.e. the noise level). At each contour level, it considers all contiguous set of pixels whose value is above that contour level. For each set it proceeds as follows:

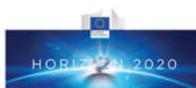

*This project is being funded by the European Union's Horizon 2020 research and innovation actions (RIA) programme under the grant agreement No 676036.*

(i) If the set does not belong to any clump, then all the pixels in that set are assigned to new a clump;

(ii) If some of the pixels in the set have been already assigned to a clump, then either (1) they all belong to one clump or (2) they belong to different (more than one) clumps;

(iii) If the (1) is true, then all the pixels of the set will be assigned to the same clump. If (2) is true then the remaining pixels (those which are not assigned to any clump yet) are shared between the existing clumps of the set using a "Friends-Of-Friends" (FOF) algorithm, i.e. the remaining pixels are assigned to existing clumps based on their distance from the clump. To find the distance of a specified pixel from a clump, the closest pixel of the clump to that pixel is considered and not necessarily the location of the clump peak.

One can summarize the CLUMPFIND algorithm as follows: Contour the data and at each level label the isolated contours as new clumps, then extend the previously identified clumps with contours at that level. For contours that surround more than one clump (blended clumps) apply the FOF algorithm. The CLUMPFIND algorithm is shown schematically in Figure 10.

One major caveat of the algorithm is its sensitivity to the adopted value for the contour interval, i.e. choosing an optimal value for the contour interval is critical. A large value for the contour interval can lead to different clumps being designated as one, whereas a small value can lead to noise spikes being mistakenly designated as real peaks. This is especially important in crowded fields where there are many blended clumps.

CLUMPFIND decomposes the data into disjoint clumps and it is easy to implement. An implantation of the algorithm in Interactive Data Language (IDL) can be found here[3]. CLUMPFIND is also available as a part of the CUPID (Section 3.2.1).

---

3    http://ascl.net/1107.014

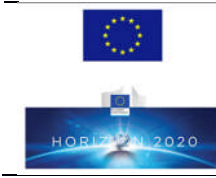

*This project is being funded by the European Union's Horizon 2020 research and innovation actions (RIA) programme under the grant agreement No 676036.*

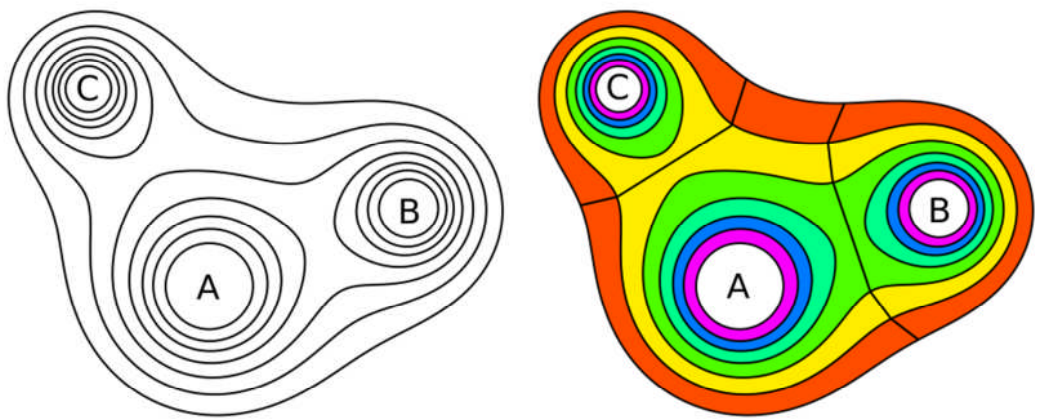

Figure 10   (Left:) sample contour lines and clumps to demonstrate how CLUMPFIND works. (Right:) clumps identified by CLUMPFIND.

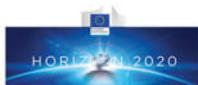
*This project is being funded by the European Union's Horizon 2020 research and innovation actions (RIA) programme under the grant agreement No 676036.*

### 3.1.1.2. FELLWALKER

This algorithm (also known as "hillwalker") is a gradient based method to identify clumps and can be applied to both 2D and 3D data.

It starts by designating all the pixels which are above a certain threshold, i.e. the background noise level. All the pixels designated as such are considered as a potential starting point and their information is stored in an array. The algorithm then proceeds by starting from the first pixel of the array and walking uphill (towards larger values of emission, density, etc.) on a line of steepest gradient until a local maximum (peak) is reached. To check if the local maximum is in fact a significant peak and to reduce the effect of statistical pixel-to-pixel fluctuations (in e.g. intensity), the algorithm searches a larger neighbourhood to look for any other pixel whose value is larger than the local peak. If such a pixel is found, the algorithm jumps to that pixel and continues uphill. If not, the peak is designated as a new clump and all the traversed pixels will be assigned to that clump. The algorithm then moves to another starting point in the array.

In the process of a finding a new clump, if the algorithm reaches a pixel which has been visited before, it stops at that point and assigns all the traversed pixels to the previously identified clump. The algorithm terminates when all the potential starting points are exhausted. Once all the clumps are identified, the algorithm attempts to merge clumps which are separated by shallow valleys. This is done to ensure that the clumps with a wide flat summit (which might have been mistakenly split into distinct clumps due to noise spikes) are re-merged and identified as one clump. This is referred to as the merging process. To reduce the effect of noise, the algorithm then makes the boundaries between adjacent clumps smooth. This is referred to as the cleaning process. Figure 11 and Figure 12 demonstrate the FELLWALKER algorithm.

Since the FELLWALKER algorithm does not use contours, it does not suffer from the issues which exist for CLUMPFIND, i.e. sensitivity to the adopted contour interval. However, there are still two minor problems with this algorithm as noted by Berry (2015) which are:

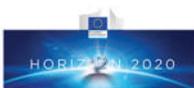

*This project is being funded by the European Union's Horizon 2020 research and innovation actions (RIA) programme under the grant agreement No 676036.*

◆ When the algorithm reaches a local maximum, it searches the neighbourhood for another pixel with a higher value to check if the local peak is a significant peak or not. If the algorithm finds such a pixel it jumps to that pixel. However, it does not check whether the two pixels belong to the same clump. Thus, the algorithm might jump from one clump to another and mistakenly merges distinct clumps.

◆ In some cases, such as "dog-boned" shaped clumps, the cleaning process leads to clumps being split into dis-contiguous parts.

The FELLWALKER algorithm is implemented as a part of the CUPID (Section 3.2.1).

### 3.1.1.3. GAUSSCLUMPS

This algorithm assumes that the data consists of many clumps all with a Gaussian-shaped density distribution. It starts by finding the brightest peak in the data and fitting a Gaussian profile to the peak, using spatial and velocity coordinates. It then subtracts the fitted Gaussian profile from the data iteratively, i.e. it proceeds by fitting a new Gaussian profile to the brightest peak in the residual map. Each fitted profile corresponds to a clump in the map. The procedure terminates when the integrated intensity of all fitted profiles becomes equal to the total integrated intensity of the observed map.

The main difference between this algorithm and other methods such as CLUMPFIND or FELLWALKER is that pixels in this algorithm do not belong to a single clump. More precisely, each pixel contributes to more than one clump and clumps can overlap. In the CLUMPFIND and FELLWALKER algorithms, however, the clumps are disjointed.

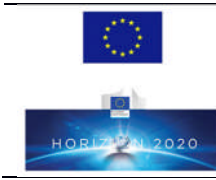

*This project is being funded by the European Union's Horizon 2020 research and innovation actions (RIA) programme under the grant agreement No 676036.*

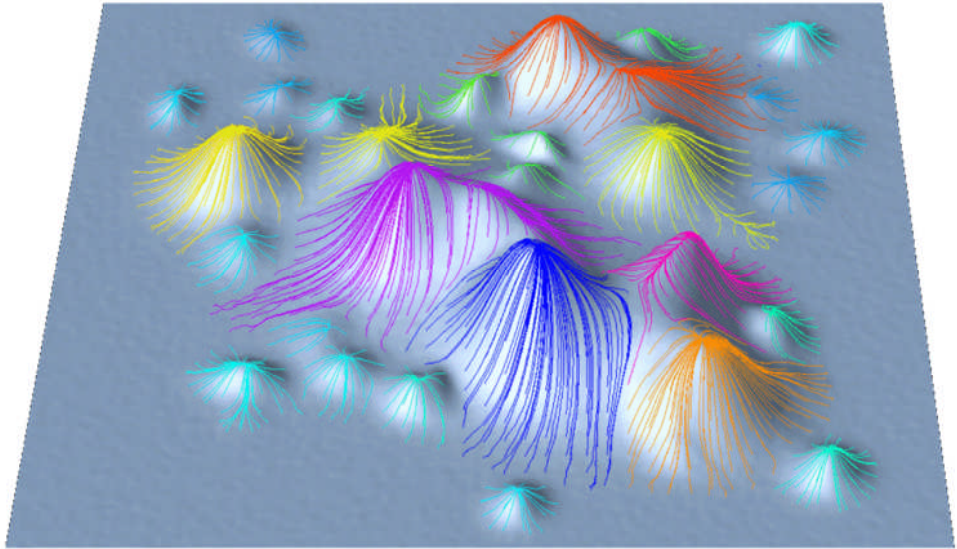

Figure 11   Schematic demonstration of the FELLWALKER algorithm. Uphill tracks (denoted by different colours) which are terminated at the same peak belong to the same clump. [4]

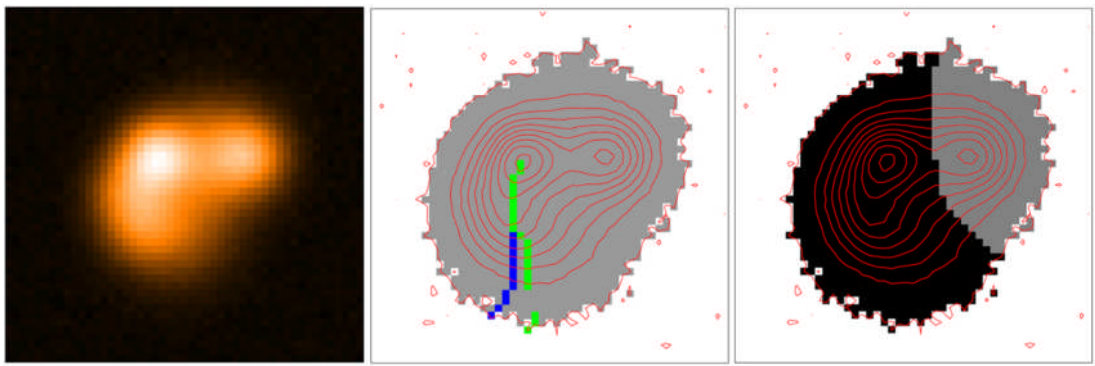

Figure 12    FELLWALKER algorithm applied to simulated 2D data of two blended clumps. (Left:) 50X50 array of simulated data. (Middle:) Contours as well as the two walks taken by the algorithm to reach the left peak. The blue walk is terminated when reaching the point which has been already traversed by the green walk. The green walk has a gap at the lowest contour level. This is due to the fact that the algorithm has reached a local non-significant peak (noise spike) and then has searched the neighbourhood of the local peak for another pixel with a larger value and has jumped to the next contour level. (Right:) The two clumps identified by the algorithm. The right clump is masked by grey. [4]

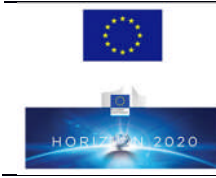

*This project is being funded by the European Union's Horizon 2020 research and innovation actions (RIA) programme under the grant agreement No 676036.*

GAUSSCLUMPS has been tested on synthetic clumpy clouds with a known power-law mass spectrum and it has been able to recover the power-law index of the mass spectrum within a certainty of ±0.1 (Stutzki and Guesten 1990). However, one caveat of this algorithm is the assumption of Gaussian profiles which is arbitrary and does not necessarily capture the real profile of clumps in the data. Moreover, the algorithm is not very effective in distinguishing individual clumps in very dense (clumpy) regions, i.e. in such regions low-mass clumps are more likely to be designated as one big clump. This can lead to an underestimation of the power-law index of the underlying mass spectrum of the clumps (Stutzki and Guesten 1990).

An implementation of the GAUSSCLUMPS algorithm can be found here[5]. Moreover, CUPID (Section 3.2.1) includes a re-implantation of this algorithm.

### 3.1.1.4. REINHOLD

The REINHOLD algorithm was developed by Kim Reinhold at the Joint Astronomy Center in Hawaii as a clump finding algorithm and comprises four main steps: (1) identifying the edges of clumps, (2) cleaning the edges of clumps, (3) filling the clump edges and (4) cleaning the filled clumps.

In the first step, the algorithm starts by assuming a set of 1D profiles running through the data. For each profile, it identifies the pixel with the highest value. If the value of the pixel is below a threshold it means there are not any significant peaks in the assumed 1D profile and the algorithm proceeds to the next profile in the set. If the value is above the threshold, the algorithm walks away from the peak in both directions along the 1D profile to find the peak ends, i.e. until it reaches the background level or it reaches a local minimum (the local gradient is below a certain threshold). REINHOLD then marks these points as the peak edges along the specified profile and proceeds to the next profile. The 1D profiles are chosen in way that they cover all pixel axes and all possible diagonal directions. Following this algorithm, one ends up with a set of rings in 2D or shells in 3D which mask the edges of peaks.

5    http://ascl.net/1406.018

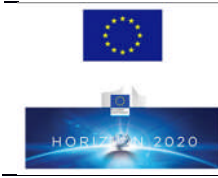

*This project is being funded by the European Union's Horizon 2020 research and innovation actions (RIA) programme under the grant agreement No 676036.*

However, the edges found this way can be noisy, e.g. there could be holes in them. To address this problem, REINHOLD applies two sequences of cellular automata to the edges, one sequence to dilate (thicken) them and another to erode (thin) them. Next, the algorithm fills the regions enclosed by the peak edges, i.e. it associates clump identifiers to pixels surrounded by the peak edges. The cellular automata applied in the second step is not 100% effective against removing holes in peak edges. This can lead to leaks during the clump filling process if there are any holes left in the peak edges. As the final step, the algorithm invokes another cellular automaton to clean up the artefacts due to the leaks. This is done by replacing each clump identifier with the most common clump identifier within its neighbourhood, which is a 3×3×3 cube of pixels in 3D and a 3×3 square in 2D.

Reinhold is available as a part of the CUPID (Sec. 3.2.1) and further details on its implementation can be found here[6].

## 3.1.1.5. GETSOURCES

Developed by Men'shchikov et al. (2012b), GETSOURCES is a sophisticated source extraction method as it is both a multi-scale and multi-wavelength algorithm. Being tested on *Herschel* images as well as simulated data, it has been shown that the GETSOURCES algorithm leads to a more reliable and complete source detection compared to the above discussed algorithms.

The core of the algorithm is (1) decomposing images at each wavelength into different spatial scales which only differ by ~5%, i.e. turning images into single-scale images; (2) cleaning images for noise and background; and (3) combining all the single-scale images across the whole range of available wavebands to benefit from the resolution information available at each waveband.

Decomposing images into single-scale images is done by convolving images with a circular Gaussian profile and subtracting them successively (known as *successive unsharp masking*), i.e.

$$I_{\lambda Dj} = G_{j-1} I_{\lambda D} - G_j I_{\lambda D} \quad , \quad j \in \{1, 2, \ldots, N_S\}$$

---

6    http://starlink.eao.hawaii.edu/devdocs/sun255.htx/sun255.htmlxref_

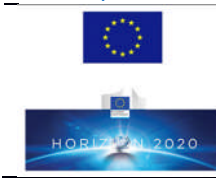

*This project is being funded by the European Union's Horizon 2020 research and innovation actions (RIA) programme under the grant agreement No 676036.*

Where $N_S$ is the number of scales, $I_{\lambda D}$ is the so-called detection image at wavelength $\lambda$, and $I_{\lambda Dj}$ is the ith single-scale decomposition of the detection image. $G_{js}$ are the set of circular Gaussians whose Full Width at Half Maximum (FWHM; denoted by S) increase, at each scale, by a scaling factor f, i.e. $S_j = f \times S_{j-1}$ and $G_0$ is a 2D delta function and the value of f is larger than unity. The value of $S_i$ varies from pixel size up to the maximum considered spatial scale. For a demonstration of successive unsharp masking refer to Fig. 2 of Men'shchikov et al. (2012b)

To clean the single-scale images from noise and background, GETSOURCES masks out pixels which are above a cut-off threshold ($\omega_{\lambda j}$) and then calculates the cut-off threshold iteratively. More precisely, the value of $\omega_{\lambda j}$ is proportional to the standard deviation ($\sigma_{\lambda j}$) over the entire image, i.e.

$$\omega_{\lambda j} = n_{\lambda j} \sigma_{\lambda j}$$

Where $n_{\lambda j}$ is a variable and is determined at each spatial scale j. The criterion based on which the pixels are masked out is $|I_{\lambda j}| \geq \omega_{\lambda j}$. The algorithm then proceeds with the remaining pixels, i.e. calculating a new value for the cut-off threshold until convergence is reached ($\Delta \omega_{\lambda j} < 1\%$). At the end of the procedure, all pixels below the converged cut-off threshold will be zeroed. This leads to a clean single-scale image which will be used later. One important aspect of the cleaning process is the adopted value of $n_{\lambda j}$ which is initially set to 6. This value has been obtained empirically and is an optimal value, i.e. smaller values than 6 leads to very low values for $\omega_{\lambda j}$ which causes images not to be free from noise, whereas larger values are not deep enough which means some of the faint sources will be missing from the results.

Once the images are decomposed in many spatial scales and are cleaned of noise and background and are also combined in wavelengths, they are considered for source extraction. The main idea of source extraction in GETSOURCES is to follow the evolution

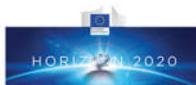

*This project is being funded by the European Union's Horizon 2020 research and innovation actions (RIA) programme under the grant agreement No 676036.*

of sources (their size and intensity) in different single-scale images as well as different wavelengths (See Fig. 8 of Men'shchikov et al. (2012b)).

To summarise one can state that the GETSOURCES algorithm has the following benefits compared to other algorithms:

◆ It effectively avoids false peaks in large and extended structures and re moves irrelevant spatial scales.

◆ Extracted sources do not need to be cross-matched with another catalogue

◆ Background subtraction

◆ Capable of deblending sources and handling hierarchical structures

◆ It is more robust and less subjective than other methods, i.e. there are not any free parameters and the algorithm can be run fully automated[7].

Despite all the above mentioned benefits, GETSOURCES could be much slower and needs more storage space than other algorithms. The exact runtime of GETSOURCES depends on several factors: the size of the images (number of pixels), number of wavelengths and spatial scales as well as the number of iterations. It can vary from a few minutes to about a week, depending on the available computational power and resources. Moreover, the intermediate images which are generated by the algorithm can occupy several hundreds of MBs to a few tens of GBs. This might not be necessarily a disadvantage, however, it makes GETSOURCES different in way that it is not a real-time processing software, i.e. one needs to prepare the images first and retrieve the source extractions results some time (typically few days) later.

To get the GETSOURCES software the reader is referred to this link[8].

---

7      For example, compare GETSOURCES with CLUMPFIND in which the results strongly depend on the adopted value for the contour intervals.

8      http://ascl.net/1507.014

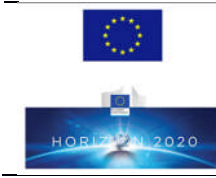

*This project is being funded by the European Union's Horizon 2020 research and innovation actions (RIA) programme under the grant agreement No 676036.*

### 3.1.1.6. CUTEX

CUTEX (CUrvature Thresholding EXtractor) is designed to detect compact and point-like sources in the infrared and is composed of two algorithms to extract sources and estimate their fluxes. It aims to alleviate problems that exist for detecting sources such as protostars, which are embedded in molecular clouds or their envelopes as they accrete mass.

To detect sources, the CUTEX algorithm generates a set of images in which pixels represent the second-order derivative of the original signal image. Since the pixels of the original image make a 2D discrete set, the differentiation is done along four different axes (i.e. columns, rows and two diagonals) using the Lagrangian differentiation method. To calculate the second-order derivative at each pixel, five consecutive points are considered to remove the effect of noise peaks and glitches.

The second-order derivative essentially determines the "curvature" of the intensity distribution at each pixel. By generating the curvature images, the positions of local maxima are found which can later be used to designate sources. In particular, CUTEX makes a mask image in which the curvature of pixels along all four directions exceed a specific threshold. Such pixels are referred to as Curvature Mask Pixels (CMP). CMPs are then grouped together if they are contiguous and make clusters. Within each of the clusters, pixels which are one sigma above the curvature value of all the surrounding CMPs are considered as statically significant maxima and are identified as candidate sources.

To estimate the flux of each source, the algorithm then fits a 2D Gaussian profile to each source, for which the free parameters are the orientation, amplitude and the FWHM of the Gaussian profile. In the fitting process, CUTEX also adds a planar background to the Gaussian profile with variable inclination and inclination direction to remove the effect of background. For blended sources, CUTEX simultaneously fits multiple Gaussian profiles to the sources.

CUTEX is effective at detecting sources in crowded regions with highly variable

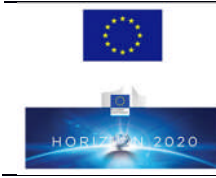

*This project is being funded by the European Union's Horizon 2020 research and innovation actions (RIA) programme under the grant agreement No 676036.*

fore/background. However, the method only works for compact (point-like) sources. The code is available in both IDL and the GNU Data Language (GDL) and can be downloaded here[9].

## 3.1.2. Filament Identification

### 3.1.2.1. GETFILAMENTS

This algorithm is based on the same approach as GETSOURCES, i.e. it decomposes the observed images into many single-scale images at different wavelengths and then extracts the filaments. The main idea behind GETFILAMENTS is that filaments are much more elongated than sources. Moreover, they occupy a larger area in single-scale images than non-filamentary structures. By thresholding single-scale images at a specific cut-off threshold, they can be filtered out. It has been shown that $\omega_{\lambda j} = \sigma_{\lambda j}$ is a good choice for detecting filaments. (For the definition of $\sigma_{\lambda j}$ and $\omega_{\lambda j}$ see Section 3.1.1.5).

In particular, the GETFILAMENTS algorithm has the following features:

◆ Detects filaments

◆ Specifies the intrinsic intensity destitution of filaments

◆ Specifies the lines of connected pixels tracing filaments crests (aka. Skeletons) at a single scale as well as the accumulated skeletons across all wavebands for the combined images

This algorithm has been incorporated in the GETSOURCES computer code. Thus, GETSOURCES effectively extracts both filaments and sources. For a demonstration of GETFILAMENTS the reader is referred to Fig. B.1 of Men'shchikov et al. (2013).

---

9    http://herschel.asdc.asi.it/index.php?page=cutex.html

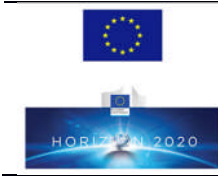

*This project is being funded by the European Union's Horizon 2020 research and innovation actions (RIA) programme under the grant agreement No 676036.*

### 3.1.3. Hierarchical Structures

As discussed in Section 2.1, dendrograms (a.k.a tree diagrams), are one of the useful tools with which discrete distributions can be studied. This is also the case for continuous distributions. Dendrograms have been shown to be effective in determining the properties of small and large structures as well as retaining merging/bifurcating information of sources. The main idea of using dendrograms to study hierarchical structures is thresholding (contouring) the data at several decreasing levels and then follow the isosurfaces of connected pixels at each level from the highest contour levels to the lowest. This technique can be applied to 2D as well as 3D data. Figure 13 illustrates a schematic representation of how dendrograms work in this context.

In addition to dendrograms, one can also use GETSOURCES/ GETFILAMENTS to study hierarchical structures.

## 3.2. Computer Programs

Many of the algorithms that have been discussed in Section 3 are available as a computer code whose download links are given in the corresponding sections. In addition, there are two more computer programs that are worth mentioning here, namely CUPID and SExtractor.

### 3.2.1. CUPID

CUPID (ClUmP IDentification and analysis package) is a set of tools to identify and analyse clumps in 1D, 2D as well as 3D data arrays. CUPID is a part of the UK Starlink project software collection which has been developed by the Starlink project[10] (until 2005), Joint Astronomy Center (until 2015) and recently by the Eastern Asian observatory[11]. CUPID is distributed under the GNU General Public License meaning that it is a free and open-source software. It is written in C and is well documented. It can be compiled from source or its binaries can be downloaded for 32 and 64-bit systems. One can refer to the CUPID official website[12] to obtain the software and the documentation.

---

10  http://starlink.eao.hawaii.edu/starlink
11  http://www.eaobservatory.org/
12  http://starlink.eao.hawaii.edu/starlink/CUPID

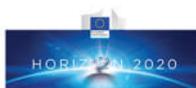

*This project is being funded by the European Union's Horizon 2020 research and innovation actions (RIA) programme under the grant agreement No 676036.*

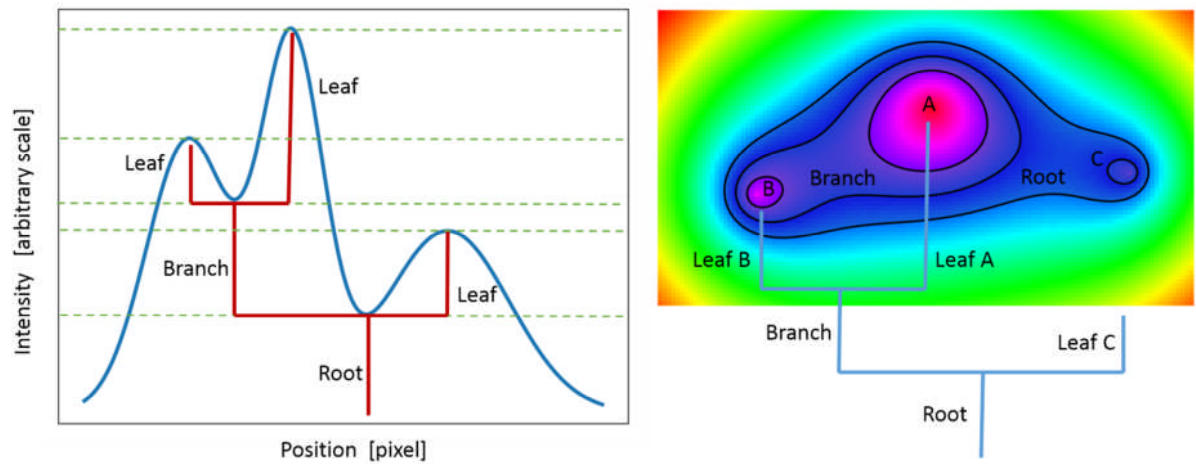

Figure 13  Schematic representation of how dendrograms are used in studying hierarchical structures. Left: The 1D intensity profile of three local peaks. Right: The 2D profile of a system containing three local peaks. For both systems, the intensity is thresholded at different levels to establish the tree structure.

The main features of CUPID are:

◆  Min-max background estimation and background removal

◆  A set of algorithms to identify clumps within emission maps. The implemented algorithms are:

1. CLUMPFIND
2. FELLWALKER
3. GAUSSCLUMPS
4. REINHOLD

◆  Providing a list of clumps as well as their properties

◆  Providing cut-out images of individual clumps

◆  A tool to create simulated Gaussian clumps with noise.

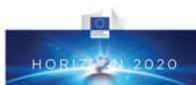

*This project is being funded by the European Union's Horizon 2020 research and innovation actions (RIA) programme under the grant agreement No 676036.*

In addition, using CUPID one can compare different clump identification algorithms.

CUPID entails a set of commands which can be easily called via the command-line, each accepting a set of arguments. The following example demonstrates the usage of the findclumps command:

```
findclumps in=INPUT_DATA out=OUTPUT_DATA method=fellwalker %
```

Where INPUT_DATA is a 1, 2 or 3D NDF (Extensible N-Dimensional Data Format[13]) file and OUTPUT_DATA is an NDF file with the same size and shape as the input file. The last parameter specifies the algorithm to identify the clumps, i.e. FELLWALKER in this example.

Another useful command within the CUPID package is makeclumps which creates 1, 2 or 3D NDF files with simulated clumps in the presence of background noise. The generated clumps have a Gaussian profile whose shape parameters can be adjusted as desired.

## 3.2.2. SEXTRACTOR

SExtractor (Source Extractor) is a computer code to detect, deblend and classify sources in astronomical images. It is a robust code which can deal with several square degrees of data and large data files with a reasonable speed. Moreover, it has the ability to estimate the total magnitude of the sources. Since its conception in 1996, SExtractor has been extensively used in the community and has been actively developed, with its latest version being released in May 2014. SExtractor has been written in C and its source code is available here[14].

SExtractor starts by background estimation and subtraction. To estimate the background, it iteratively clips the local background histogram until convergence is reached at ±3σ around the median. The detection mechanism of SExtractor is based on thresholding. As a part of the detection process, it convolves images with a convolution mask. Depending on the

---

13    http://starlink.eao.hawaii.edu/docs/sun33.htx/sun33.html
14    http://ascl.net/1010.064

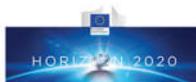

*This project is being funded by the European Union's Horizon 2020 research and innovation actions (RIA) programme under the grant agreement No 676036.*

shape of the sources and the fields in which they reside, different convolution masks can be used with SExtractor. For example, for stars (point-like sources) a Point Spread Function (PSF) can be used.

A major feature of SExtractor is its ability to deblend sources by applying a threshold criterion as follows. The intensity of contiguous pixels for each extracted source is contoured into 30 levels equally spaced on a logarithmic scale between the detection threshold (minimum) and the peak value (maximum). Then the algorithm starts from the peak maximum and goes towards the minimum and looks for junctions. At any junction $t_i$, a branch will be considered as a new (separate) source if the integrated pixel intensity above $t_i$ exceeds a specific fraction of the total intensity of the composite object.

SExtractor is specifically tuned to discriminate reliably between stars and galaxies in large extragalactic surveys, which is achieved via a neural network trained by artificial images. Moreover, it is able to detect sources in star fields which are not highly crowded.

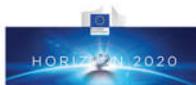

*This project is being funded by the European Union's Horizon 2020 research and innovation actions (RIA) programme under the grant agreement No 676036.*

# 4. Clustering in Astrophysics

Grouping the data in clusters and performing statistical analysis on data sets are important tasks in science and hence in astrophysics. Developing new clustering algorithms are an important and dedicated part of Applied Mathematics and Statistics, Computer Science and Knowledge Database Discovery (KDD) fields. Initially, most of the tools implemented in the field of astrophysics come from that expert knowledge domains, and were first applied in extragalactic field. These works allowed the first characterization of spatial and kinematic cosmic web using statistical tools such as correlation function, nearest neighbour statistics and clustering algorithms as early as in the 1930's (for an historical review see Peebles, 2001). That was initiated half a century before these techniques were applied on stellar data sets (first publication early 1990's, Gomez et al., 1993). With the impulse of the era of big data in Astronomy, the thematic subfield "Astrostatistics and Astroinformatics" is currently under great development to establish a dedicated bridge between statistical and data mining tools and astronomical data (Feigelson, 2016 and Brescia and Longo, 2013; for an historical perspective Feigelson, 2009 and see the Astrostatistics and Astroinformatics Portal[15]).

Below we focus on those works aiming at characterizing the distribution of stars and gas by identifying their spatial substructures and/or their velocity components, as these are the most relevant to the goals of the SFM project (Section 5). In addition we briefly review the promising tools used and developed in extragalactic field that can be interesting to investigate and apply within the framework of the SFM project. For a full overview of statistical clustering and statistical methods applied to astrophysics (which is outside the scope of this report) the reader is referred to Feigelson and Babu (2012) ; Murtagh and Contreras, (2012); Murtagh and Heck, (1987); Starck and Murtagh, (2006).

---

15    https://asaip.psu.edu

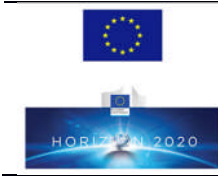

*This project is being funded by the European Union's Horizon 2020 research and innovation actions (RIA) programme under the grant agreement No 676036.*

## 4.1. Stars, cluster of stars and clustering

### 4.1.1. Spatial Studies of Star Clusters & Associations

Two types of works related to the analysis of spatial distributions may be distinguished. The first focuses on spatial analysis in a global sense, i.e. it aims at the characterization of the distribution of points set as a whole. It is the case, for example, when the correlation function is used to evaluate the degree and the regimes of clustering.

The second type of work aims at extracting (sub-)structures as topological entities to further characterize them and derive their geometrical and physical properties. It's in this latter case that clustering algorithms are mandatory.

These two types of studies developed in the field of stellar astrophysics were/are inspired from the works done in the extragalactic field. The latter came out as early as in the thirties and were applied to characterize the distribution of galaxies and cluster of galaxies (see the historical perspective done on the origin of the correlation function in astronomy by Peebles, 2001). The first work to derive the correlation function from a distribution of stars was performed almost 25 years ago by Gomez et al. (1993) on the young stars within Taurus, the nearest low-mass star forming complex located at 145 pc from us. Thus, extragalactic field has a clear advance –half a century- in the statistical and clustering techniques they use to analyse the cosmic web.

Naturally, the two types of studies (i.e. global and topological) are complementary and even sometimes the same tool, such as the MST, may serve at the two perspectives. Indeed, the MST method was applied for the first time in the extragalactic astrophysics by Barrow et al. (1985) to analyse the spatial distribution of galaxies in the cosmic web (1) through the MST branches distribution statistics and (2) the identification of the individual topological structures, once applied filtering procedures on the MST such as nodes pruning and edges removal. In the context of young star clusters, Gutermuth et al. (2009) proposed a procedure to identify a length threshold above which the edges may be removed in order to identify the clusters i.e. the remaining connected sub-trees. This was in turn specifically

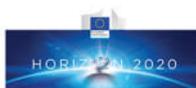

*This project is being funded by the European Union's Horizon 2020 research and innovation actions (RIA) programme under the grant agreement No 676036.*

used in four star forming regions (Taurus, Lupus3, ChaI, and IC348) to study the properties of 14 groups of stars altogether (Kirk and Myers, 2011).

The MST was also used to define useful parameters to assess the degree of substructuring. Wright et al. (2014) found considerable substructure signature for an X-ray selected sample of young stars the nearby young association Cygnus OB2 by using the Q parameter (Cartwright and Whitworth, 2004) which is defined as the ratio of the average branches length of the MST over the mean separation of all pairs of stars. The same parameter Q was widely used from that point on in different works to assess the degree of clustering as, for example, in the work of (Schmeja et al., 2008) performed in Serpens, Ophiucius and Perseus star forming regions or to assess the dynamical status of a star cluster in numerical simulations (Parker et al., 2014). Another important issue in the spatial study of the stars is to assess the distribution of massive stars with respect to the low-mass ones. This can be done using e.g. the m-Σ stellar surface density methods or amongst others, the Λ method (Allison et al., 2009) that uses MST tree of sub samples of massive and low-mass stars. A comparison and discussion on different methods assessing global measure of the substructuring degree and the mass segregation can be found in Schmeja (2011), Küpper et al. (2011) and (Parker and Goodwin, 2015).

More recently a new parameter, the angular dispersion, was proposed by (Da Rio et al., 2014) to evaluate the substructuring degree of the Orion Nebula Cluster. This was then applied (Jaehnig et al., 2015) to the 15 clusters included in the Massive Young Star-forming compleX (MYStIX) (Feigelson et al., 2013) to study the effect of dynamical evolution. Very recently, new parameters have been proposed assess the degree of fractality of the star cluster spatial distribution (Jaffa et al., 2017). On the same MYStIX clusters sample, Kuhn et al. (2014) used a 2D-parametric components mixture model. The six free parameters of the surface density model (isothermal ellipsoids: flat core + power-law wings) of each subcluster are derived by maximum likelihood optimisation. Using the Akaike Information Criterion (AIC) the number of clusters (statistical model selection) is derived based on the comparison of their relative quality measure for the given set of data.

Caballero and Dinis (2009) applied the DBSCAN algorithm to the Hipparcos stars catalogue to identify aggregates in large stellar complexes, OB associations, and young open clusters.

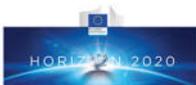

*This project is being funded by the European Union's Horizon 2020 research and innovation actions (RIA) programme under the grant agreement No 676036.*

The spatial discrete distribution of stars may also be turned into a continuous density map by the convolution of the spatial point distribution with a kernel function, in such a way that a stellar density map is obtained (Silverman, 1986, primarily used in the star association framework by Alfaro et al., 1992 and Gomez et al., 1993). This map can then be analysed the same way as the extinction/emission gas map from molecular clouds (see Section 3 for the description of these techniques). From those continuous data an agglomerative hierarchical clustering may be applied on the nested isosurfaces in a two dimensional extinction map or three dimensional molecular line data cube (i.e. position-position-velocity data cubes) to obtain dendrograms as described in Rosolowsky et al. (2008) . It was first proposed as a general idea by Houlahan and Scalo (1992) and recently applied in star cluster analysis by Gouliermis et al. (2010, 2015).

## 4.1.2. Spatial & Kinematic Studies of Star Clusters/Associations

In the context of the GAIA-ESO survey, one of the first works done to identify kinematics components in the young star cluster Gamma-Velorum was by Jeffries et al. (2014) who identified YSO radial velocity components using a mixture of Gaussians.

A generalization of the Λ−parameter (Allison et al., 2009) index, Λ− (RV), where RV stands for Radial Velocity, was defined in the 3D sub-space formed by two position coordinates and radial velocity of star-forming regions (Alfaro and González, 2016). As Allison et al. (2009), they used the MST to build a spectrum of the Λ− (RV) parameter per bin of radial velocity (a technique they later called SVG for Spectrum of Kinematic Grouping) to identify kinematically clumpy structures. This technique was applied to the Cygnus OB1 association and the results were compared to those obtained with the OPTICS algorithm described in Section 2.2.2 (Costado et al., 2017). Using the two techniques the authors identified two main kinematics groups which have different RV, centroids and are well spatially separated; and a third component which was only detected by the SVG technique that may be a common and older stellar component.

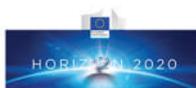

*This project is being funded by the European Union's Horizon 2020 research and innovation actions (RIA) programme under the grant agreement No 676036.*

Recently, (Vicente et al., 2016) performed the spatial-kinematic analysis of the old open Galactic cluster NGC2548 using the KDE (Kernel Density Estimation, Silverman, 1986) method to produce density maps in both the projected spatial sky and in the proper motion space. They identify under human supervision three spatially separated cores that have their own counterparts in the proper motion distribution.

## 4.1.3. Studies of Galaxy Clusters and the Cosmic Web

Galaxy redshift surveys such as the Sloan Digital Sky Survey (SDSS) and the 2dF Galaxy Redshift Survey (2dFGRS) show that galaxies and cluster of galaxies are spread out in a fairly intertwined complex structure that defines the cosmic web. This network consists of galaxies and galaxy clusters, which are interconnected through filaments, sheets and walls that themselves leave room to large regions with almost no galaxies, the voids. Although not immediately obvious, methods used and being developed for the study of large scale structure in cosmology have value for star formation as well.

As already underlined in the previous subsection, we may distinguish two types of studies those that intend to characterize globally the spatial distribution with a set of parameters or based on statistical distribution (clustering statistics, see for a review (Sahni and Coles, 1995; Bertschinger, 1998; Martinez and Saar, 2002) and those that intend to identify and extract the individual structures.

### 4.1.3.1. Clustering Statistics

In the first type of studies, the two-point correlation function naturally (and to a much lesser extent the higher-order correlation functions) is at the first place (e.g. Kerscher et al., 2000; Landy and Szalay, 1993; Peebles, 1973, 1980) since they have most extensively been used.

Some other tools based on graph theory and network science have been applied to galaxy point distribution. The early work made use of percolation statistics with a FOF procedure (Shandarin and Zeldovich, 1983; Sheth et al., 2003). The MSTs and related statistics have been extensively used in that extragalactic context (Adami and Mazure,

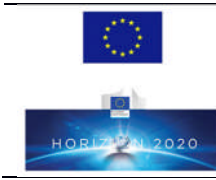

*This project is being funded by the European Union's Horizon 2020 research and innovation actions (RIA) programme under the grant agreement No 676036.*

1999; Barrow et al., 1985b; Krzewina and Saslaw, 1996; Bhavsar and Splinter, 1996). It has been also used to derive structure functions to evaluate the type of morphology, i.e. clusters, filaments or walls (Colberg, 2007 derived from the work of Babul and Starkman, 1992). More recently works use network metrics to characterize the topological structure of cosmic webs (Hong and Dey, 2015; Hong et al., 2016).

The attempts to characterize the cosmic web cover a large spectrum of techniques and audience starting from a rather confidential work done on shape statistics (Luo et al., 1996) to the widely used Minkowski functionals that estimate within a 3D Euclidean space the volume, surface and curvature of the density field (Gott et al., 1990; Mecke et al., 1994; Sahni et al., 1998; Sheth et al., 2003; Spergel et al., 2007). The Minkowski functional includes the genus statistics computed on a smoothed density field has been proposed to evaluate the degree of connectedness as a function of density enhancement (Gott et al., 1987); the genus is the topological invariant given roughly speaking the number of handles (dually i.e. the plain voids) of a one contour surface. In complex situations, it is the number of voids minus the number of isolated regions. It may be computed as the integral of the Gaussian curvature over the surface contour.

## 4.1.3.2. Structure Identification and Extraction

The identification and extraction of (cluster-type) halo structures were developed from the seventies and are mostly based on a FOF algorithm type (Section 4.1.3.3). The work performed on elongated structure (filaments) is more recent (see Section 3), and is of particular interest since it is likely to be readily applicable to similar features in star forming regions. These methods are aimed at studying the filamentary nature of the "cosmic web", such as general filament finders; e.g. González and Padilla (2010). They first identify the nodes of high density using a nearest neighbour algorithm to build a backbone of the filaments network and then assign to a particular filament galaxies and galaxy clusters that are more strongly linked (energy bound) to the filament. This is a strategy that appears particularly interesting in the context of star forming region to identify the filamentary structure of the stars pattern.

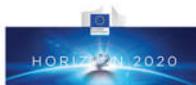

*This project is being funded by the European Union's Horizon 2020 research and innovation actions (RIA) programme under the grant agreement No 676036.*

Early works made use of MST and percolation statistics with a FOF procedure (Shandarin and Zeldovich, 1983; Sheth et al., 2003). Other works investigate the neighbourhood between two clusters of galaxies based on the empirical criteria for the parameters selections, i.e. the number of nearest neighbours inside an upper sized rectangular window (Colberg et al., 2005) or analysing the neighbourhood based on angular sections (Pimbblet, 2005).

In the framework of spatial marked point process, the Bisous model (Tempel et al., 2014), has been able to extract the filamentary network of the cosmic web. The basic hypothesis that the distribution of galaxies may be described as a random configuration of connected and aligned cylinders. Galaxies are thus supposed to be locally grouped together inside small cylinders that may combine to form a filament if the neighbouring cylinders are aligned in similar directions.

Of particular interest is the multi-scale and automatic morphological classifier tools ("NEXUS" and "NEXUS+") developed by Cautun et al. (2013) which identifies cosmic structures (filaments, walls, voids, clusters) in a scale free way, without preference for a certain size or shape based on the Hessian matrix and its eigenvalues that allow to associate to each point a geometrical status. Testing of these tools using N-body simulations has produced encouraging results, showing that Nexus+ correctly identifies the most prominent filaments and walls even for faint structures and demonstrated the tools superiority over other Hessian and topological-based methods to identify filamentary structures (for a full discussion see Section 7 of Cautun et al., 2013).

It's interesting to note that, with the exception of the MST technique, there are very few links on modern clustering algorithms developed in computer science and the identification of the cosmic web. However, density based clustering algorithms have been utilised in the detection and classification of galaxies. (Tramacere et al., 2016) has successfully used the DBSCAN algorithm to detect overdense structures (galaxies), and then applied the DENCLUE algorithm to discriminate between the spiral and elliptical galaxies based on the fact that the former have multiple local maxima as the latter have only one extrema associated to the centre of the elliptical galaxy.

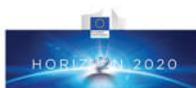

*This project is being funded by the European Union's Horizon 2020 research and innovation actions (RIA) programme under the grant agreement No 676036.*

### 4.1.3.3. Velocity and Spatial Structure Detection

A robust clustering statistics, the $\Delta$-statistics using velocity kinematics and sky projected positions, brings an important indicator to quantify the degree of substructuring in galaxy clusters (Dressler and Shectman, 1988; Knebe and Müller, 2000).

Amongst the classical statistics used to identify the space-velocity structure, Einasto et al. (2012) use 3D normal mixture modelling, the Dressler-Shectman test, the Anderson-Darling and Shapiro-Wilk tests, as well as the Anscombe-Glynn and the D'Agostino tests, to find the peculiar velocities of the main galaxies, and use principal component analysis to characterise their results.

Amongst more sophisticated methods to detect stream flows in the Cold Dark Matter (CDM) numerical simulations of the cosmic web Shandarin et al. (2012) set as a cornerstone of their method the tessellation of the 6D-phase Lagrangian space mapped on the Eulerian space.

At much lower scale, our Galaxy, other type of stream flows focus the interest of Galactic astronomers (Helmi, 2008) as they realize that a lot of spatial and kinematic substructures composed the halo of the Galaxy as probable imprints of the history of the formation of the Milky Way. Recently, Sans Fuentes et al. (2017) used the FOptics algorithm to detect tidal debris as overdensities in the phase-space in the Galactic Halo. The GAIA era, and the gain of precise astrometric and kinematics measures, provide new perspectives in order to studies the halos. Different methods have been developed to use kinematical information to detect substructure in phase-space, such as the velocity correlation function and overdensities in energy-angular momentum space (Helmi et al., 2017). Defined as gravitationally bound and substructured objects, all the works, tools and algorithms developed from the seventies offer a great toolbox that covers velocity outliers identification which allow detection of subhaloes, bound overdensities in phase space and tidal streams using a FOF algorithm (Elahi et al., 2011, 2013; Pujol et al., 2014) to estimate the density of points in phase space using a tessellation based on a binary tree (Ascasibar and Binney, 2005). All kind of tools that

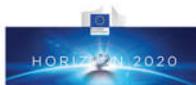

*This project is being funded by the European Union's Horizon 2020 research and innovation actions (RIA) programme under the grant agreement No 676036.*

can also easily be adapted within the framework of star cluster/association formation and evolution. The reader is referred to the extensive review and comparison of all the algorithms used to detect halos (most of them are FOF algorithms) (Knebe et al., 2013), and also Behroozi et al. (2015) who compare five halo-finding algorithms (AHF, HBT, ROCKSTAR, SUBFIND, and VELOCIRAPTOR) in the recovery of halo properties for both isolated and cosmological major mergers; their conclusion being that depending on the chosen algorithm, there are some differences on the details of the structure retrieved. This is a strong argument for the requirement that a same algorithm has to be used in order to compare the structures and statistical properties of different star forming regions.

Alternatively to these algorithms, a very interesting method exploits the fact that the if the halo is composed by the debris coming from the disruption of smaller subsystems then they should remain on a same great circle (Johnston et al., 1996; King et al., 2012; Mateu et al., 2011, 2014, 2016).

## 4.2. Molecular Clouds

The Herschel space observatory has allowed us to probe the structure of molecular clouds from deca-parsec scale (the size of an entire cloud) down to the deci-parsec scale (the size of dense cores), owing to its high sensitivity to thermal dust emission and high resolution imaging. Herschel has revealed that the cold interstellar medium (ISM) exhibits a universal filamentary structure. The prevalence and omnipresence of filaments (i.e. over-dense and elongated structures) is independent from the star-forming content of the cloud. In other words, filaments have been observed even in non-star-forming and diffuse regions. This finding implies that the filaments can play a vital role in the star formation as they form earlier than stars (André et al. 2014; André 2015).

Moreover, Herschel images have shown that in addition to the filaments, there is also a network of sub-filaments which are perpendicularly connected to the main filament, indicating that the main filaments accrete mass via the sub-filaments (André et al. 2014; André 2015).

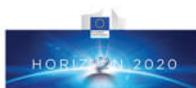

*This project is being funded by the European Union's Horizon 2020 research and innovation actions (RIA) programme under the grant agreement No 676036.*

In addition, the radial profile of filaments can be described by the following function (Arzoumanian et al. 2011)

$$\rho_P(r) = \frac{\rho_c}{\left(1 + \left(\frac{r}{R_{flat}}\right)^2\right)^{\frac{p}{2}}}$$

Where $\rho_c$ is the central density, $2 \times r_{flat} \approx 0.1 \ pc$ is the diameter of the central inner region of the filaments and appears to be almost constant for all filaments in the nearby Gould Belt clouds and $p = 2 \pm 0.5$ is the power-law exponent for large radii ($\rho_P(r)$ approaches a power-law for $r \gg r_{flat}$).

It is important to point out that before Herschel there were studies which emphasized the presence of filaments and their importance for star formation (e.g. Schneider and Elmegreen 1979, Hartmann 2002). However, the study of faint filaments and the discovery that filaments are present on all scales, became possible with the Herschel Gould Belt (André et al. 2010) and Hi-Gal (Molinari et al. 2010) surveys as well as sophisticated methods such as GETSOURCES and GETFILAMENTS (Men'shchikov 2012, 2013) which were developed to detect and study the compact and elongated sources in Herschel images.

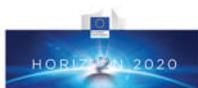

*This project is being funded by the European Union's Horizon 2020 research and innovation actions (RIA) programme under the grant agreement No 676036.*

# 5. StarFormMapper

The key aim of the StarFormMapper (SFM) project is to combine data from two of ESA's major space missions, Gaia and Herschel, together with ground based facilities, to constrain the mechanisms that underlie massive star and star cluster formation. Our scientific results will underpin the study of how all galaxies evolve.

Although low mass star formation is well understood, there is little consensus about the formation of massive stars, despite considerable recent theoretical and observational effort. Primarily this is due to the paucity of high mass stars coupled with their more rapid evolution compared to their low mass counterparts. A fundamental roadblock in our understanding of high mass evolution is that their formation occurs in associations, groups and clusters but we have only a limited understanding of how clusters and associations of high mass stars themselves form and evolve, or how the massive stars forming within them contribute to this process.

There are currently two "traditional" models of star formation which are favoured: monolithic collapse (e.g. Krumholz et al, 2009) and competitive accretion (e.g. Bonnell et al 2004). Both lead to disk accretion of material onto the central star in order to overcome issues with radiation pressure halting infall, but start from very different initial conditions. The former assumes the monolithic collapse of a single molecular gas clump in which turbulence has helped to raise the Jeans mass, whereas the latter assumes stars in clusters initially form like their low mass counterparts but funnelling of the gas to those at the centre of the gravitational potential cause them to grow larger. For the massive star formation scenario, core mergers in very dense regions is considered (see Zinnecker & Yorke 2007 for a full review on the different massive star formation models). More recently, Vasquez-Semadeni et al. (2017) proposed a very interesting model in which there is a hierarchical star cluster assembly in globally collapsing molecular clouds. There are three basic tests that can discriminate between these different models: (1) sub-clustering/structure; (2) mass segregation; and (3) the relative dynamics of the stars/gas. The aim of the SFM Project is to conduct all three of these tests, using combination of observational and simulated data. This report addresses the first; it reviews existing clustering methods to identify the optimal technique for finding and quantifying sub-clustering/structure of stars and gas.

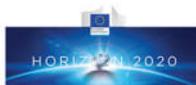

*This project is being funded by the European Union's Horizon 2020 research and innovation actions (RIA) programme under the grant agreement No 676036.*

As discussed in previous sections, there are a large number of statistical clustering techniques which are readily available but each have their weaknesses. For example, some suffer from loss of data and are sensitive to erroneous previous cluster merging (e.g. hierarchical; Section 2.1.1), some produce clusters of a particular shape (e.g. hierarchical, iterative relocation based; Section 2.1.1, 2.2.1), and nearly all require "tuning" of their input parameters which makes derivation of significance difficult.

It is generally agreed that statistically the 'best' method of detecting clusters in a data set is dependent on what what one is looking for and in what context. It is therefore essential that we identify the optimal means to do that for the astronomical data (Gaia, Herschel, simulations) and context we are considering (sites of active and/or recent massive star formation). Our goal is to adopt and adapt some of the discussed statistical techniques and develop new ones to analyse the clustering of stars and gas when both are present. In this Section we discuss challenges specific to, and the optimal existing clustering methods for, identifying sub-clustering/structure of stars and gas; and present our new statistical tools that we are currently implementing in the SFM project.

## 5.1. Statistical Methodology in the Presence of Missing Data

One of the most important problems SFM will need to overcome is missing and partial observational data sets. As discussed in Section 4 many relatively (and not so) simple statistics have been used in astronomy to identify and quantity (sub-) structure and clustering on a variety of scales, from stars and gas to the cosmic web.

Despite this however, these methods on their own are not sufficient in the context of the SFM Project due to the expected significant missing data in our studied distributions caused by observational limitations i.e. it is reasonable to expect thresholds of one or more parameters and at least partially censored photometry due to the variable (and typically high) extinction associated with star forming regions. For example, Gaia is expected to detect down to mid-G stars for all nearby star forming regions and indeed this concurs with the completeness limits found by the Leeds node for SFM candidate star formation sites (NGC2264, NGC244, NGC6618, NGC6611, Pismis 24, Cygnus X, NGC7538, W3, W48, RCW79). As such, any analyses of these regions using Gaia data sets will be 'missing' the later spectral type stars and since the

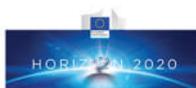

*This project is being funded by the European Union's Horizon 2020 research and innovation actions (RIA) programme under the grant agreement No 676036.*

extinction is variable across the cluster, the degree of censoring will not only be dependent on the object brightness but its location within the full 3D volume of the cluster, which obviously poses a significant problem for studies of cluster substructure, sub-clustering and mass segregation.

A mainstream statistical solution for dealing with missing objects in data sets is replace it with substituted values, a process known as data imputation. For example, bayesian techniques can be useful in deriving the underlying probability distribution of samples with incomplete data (e.g. multiple imputation using Markov Chain Monte Carlo - Schafer 1999). Unfortunately this type of data imputation is strictly wrong for the data sets of SFM, since our the missing data will not be a random sample of the complete set (i.e. for Gaia data sets only spectral types later than mid-G are missing). The combination of statistically identifying, classifying and quantifying clustering in data sets with significant (non-random) missing data is not well described in the literature. Our approach to solve the problem of missing and partial observational data sets will be to compare the observational data with the results of the star formation simulations currently being developed by the Cardiff node (see deliverables D3.1, D3.2) to statistically test and assess which approach is best.

## 5.2. New Tools

The Leeds node have developed a novel tool to quantify the degree of clustering of each particle in a discrete distribution (Figure 14). Based on the nearest neighbour method, the tool determines the parameter space distance between the star i and every other star j (j≠i) in an observed distribution, compares this to a generated spatially uniform distribution, and assigns each star a 'clustering index' (the higher the index the more clustered a star is). Promising results have been obtained for NGC2264, correctly identifying the Spike, Cone and S-Mon sub-clusters (Teixeira et al. 2006, Sung et al. 2009) and indicating the presence of a fourth sub-cluster (Buckner et al., in prep). The advantage of this tool is that it can be applied to quantify the degree of clustering of each object in a discrete distribution, in any specified parameter space without preference for size, shape or number of number of dimensions (2D, 3D, 6D). Development of the tool is ongoing and we will test this method as a means of looking for subgroups (spatially, kinematically or both) as well as examining mass segregation in clusters.

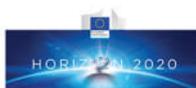

*This project is being funded by the European Union's Horizon 2020 research and innovation actions (RIA) programme under the grant agreement No 676036.*

The Grenoble node has proposed new statistics, the one-point correlation function and the mutual nearest neighbour statistics to study the clustering at the smallest scale within the star forming region. They applied these tools to the Taurus star forming complex (Figure 17) and discovered a new population of ultra-wide systems of high order multiplicity composed for the great majority of multiple/binary systems (Joncour et al., 2017). In a second part, they used the DBSCAN algorithm (Figure 15, Figure 16) and defined a procedure to detect the local spatial overdensities at a high significance (99.8%) above random expectation discovering 20 structures that contain ¾ of the youngest objects (Class I objects), probing that these structures are the NESTs (Nested Elementary STructures) for the birth and the childhood of young stars (Joncour et al, in prep). They currently work on a multiscale analysis of the spatial distribution of stars using a recursive algorithm based on DBSCAN to build a clustergraph associated to an Ntree (the analogue of dendrogram for the binary tree). They will apply that tool to characterize the structural and hierarchical property of Taurus complex. Furthermore they have determined the 3D global structure of the young Upper Scorpius OB association using the robust covariance statistical method on the GAIA parallax of the TGAS catalogue in the DR1 release (Figure 18). This catalogue was complemented by the kinematic parallax evaluated for the fainter stars. We outline the difference in the spatial distribution of the bright stars with respect to the fainter stars (Galli et al, in prep).

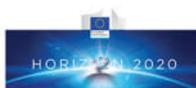

*This project is being funded by the European Union's Horizon 2020 research and innovation actions (RIA) programme under the grant agreement No 676036.*

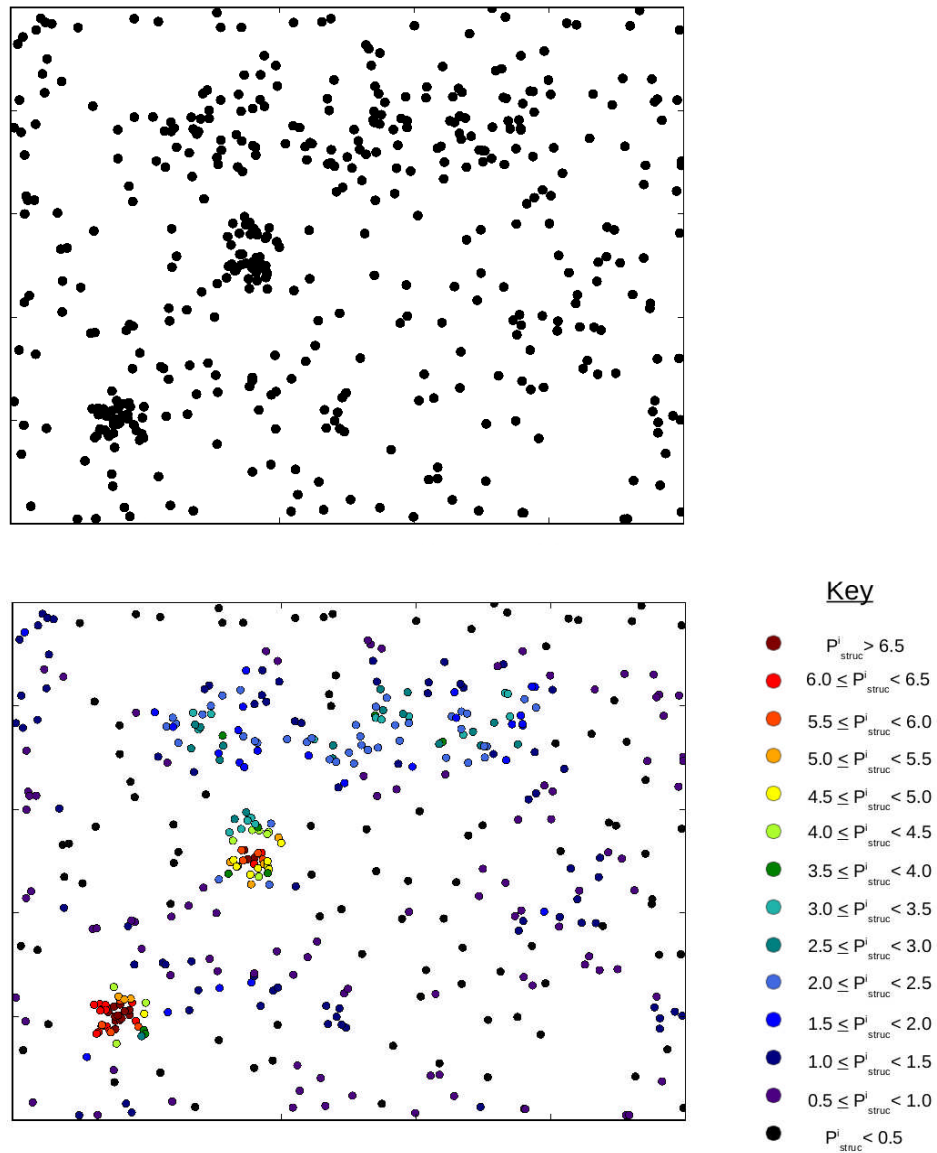

Figure 14 Demonstration of the tool developed by the Leeds node to quantify clustering. (Top:) the data set pre-application of the tool; (Bottom:) the data set post -application of the tool, with a 'clustering' index assigned to each object (see text for details). An index of $P^i_{struc} > 1.0$ indicates object is clustered (higher values indicate object is more clustered), $0 < P^i_{struc} < 1.0$ indicates object is isolated (lower values indicate object is more isolated) and $P^i_{struc} = 1.0$ indicates object is neither clustered nor isolated.

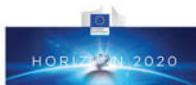

*This project is being funded by the European Union's Horizon 2020 research and innovation actions (RIA) programme under the grant agreement No 676036.*

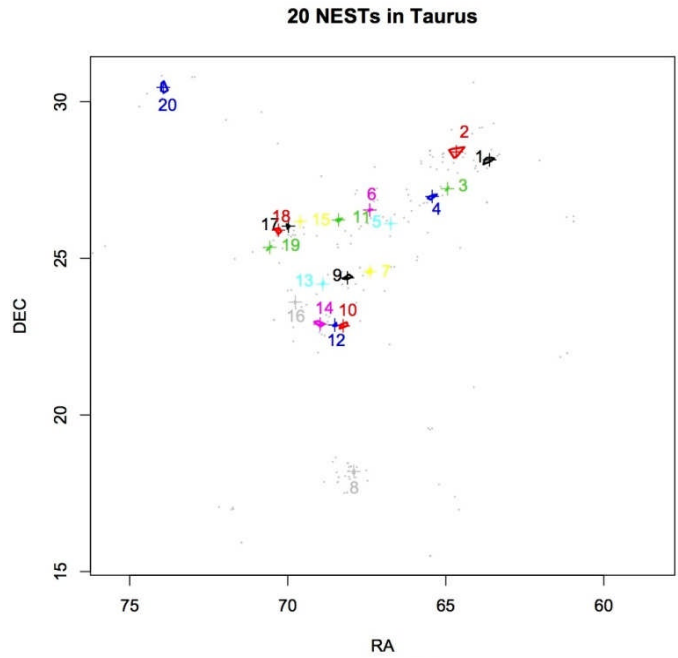

**Figure 15 Using the DBSCAN algorithm (ε=0.29pc, Nmin=4) to detect the spatial structure at the 99.8% above the mean uniform random distribution, we detect 20 NESTs in the Taurus star forming region that contain altogether nearly half the young star population and containing 75% of the youngest star objects (Class I). (image reproduced from Joncour et al, in prep)**

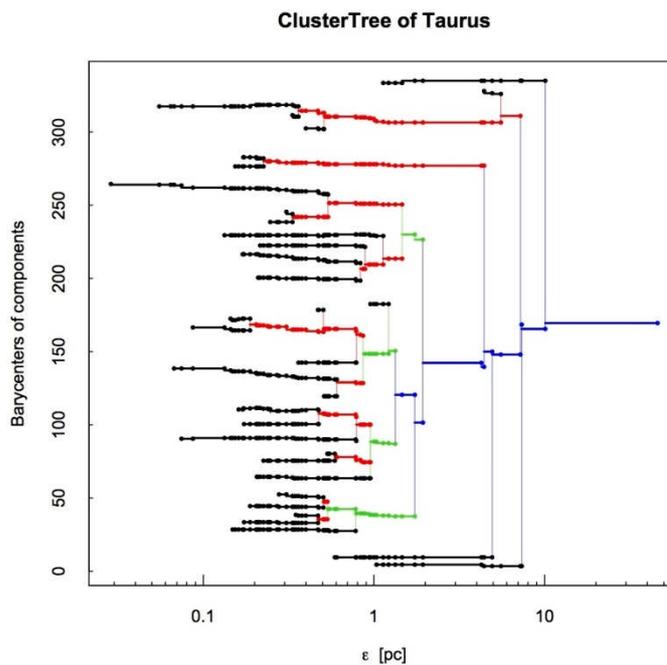

Figure 16 We implement a hierarchical DBSCAN algorithm and build a clustergram (ClusterTree) as a hierarchical skeleton of the spatial structure of the young stars in Taurus. Colours indicate the Strahler order: 1 (black), 2 (red), 3 (green), 4 (blue). (image reproduced from Joncour et al., in prep)

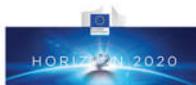

*This project is being funded by the European Union's Horizon 2020 research and innovation actions (RIA) programme under the grant agreement No 676036.*

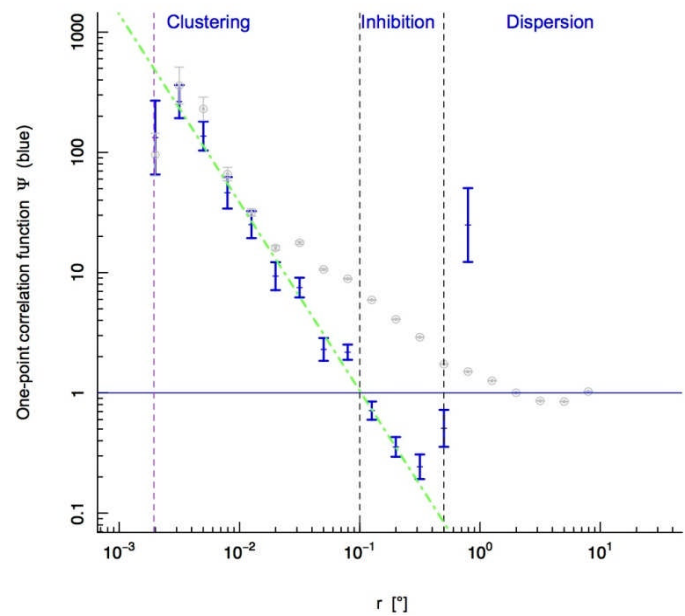

Figure 17 Defining the one-point correlation function as a new statistical function to complement the two point correlation function, we discover a new multiplicity regime going up 60 kAU. (image reproduced from Joncour et al, 2017).

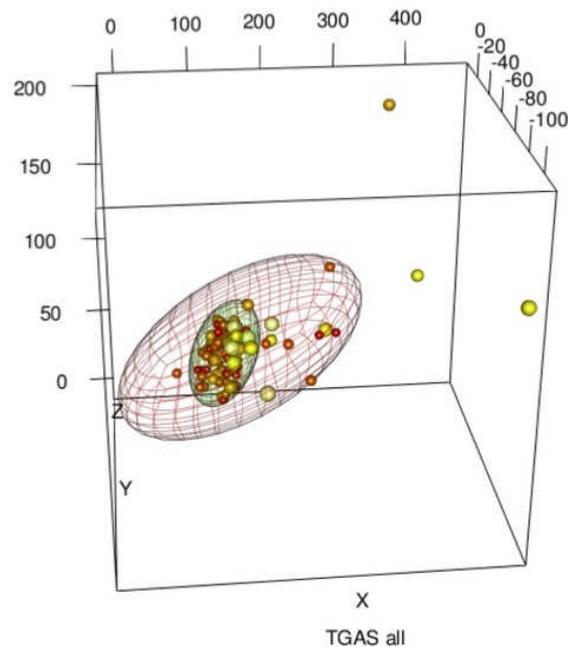

Figure 18 Using the GAIA parallax (DR1 release) of the stars in the young association of OB stars Upper Scorpius, we compute the geometrical characteristics of its 3D main spatial structure using robust location and scatter estimators for 3D spatial analysis.

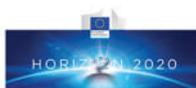

*This project is being funded by the European Union's Horizon 2020 research and innovation actions (RIA) programme under the grant agreement No 676036.*

# 6. Summary & Conclusions

The StarFormMapper project will combine data from two of ESA's major space missions, Gaia and Herschel, together with ground based facilities and simulations, to constrain the mechanisms that underlie massive star and star cluster formation. Specifically we aim to discriminate between the "traditional" models of star formation by conducting tests for (1) sub-clustering/substructure, (2) mass segregation, and (3) the relative dynamics of the stars/gas.

This report has addressed the first; it is a review of existing clustering tools for both discrete (stellar) distributions and continuous (gas) distributions, in order to identify the optimal tools that are suited for our specific study that focus on the study on the massive star formation processes via their 2D/3D spatial and 2D/3D kinematic imprints they left in the gas and stellar distribution.

After a brief introduction in Section 1, we made a very general (not specific to astronomical data) review of the main clustering approaches for discrete distributions in Section 2, from the most traditional techniques (hierarchical and K-means approaches) to the more recent ones (grid and density based approaches). These techniques were developed in computer science field but are widely used in all fields from the Social Sciences to Biology and Physics. In Section 3 we made a more specialist (specific to analysing astronomical gas data) review of the main clustering approaches for continuous distributions.

For the purpose of this report we made a rather arbitrary distinction between what we call "discrete" and "continuous" distributions. The former is associated to point-like distributions, whereas the latter is associated to image data. Our distinction is arbitrary because digital images may also be considered as a collection of points (pixels) that can be grouped together based on attributes (temperature, brightness, etc.) the same way as our definition of discrete distributions. Nevertheless there is a kind of "phase transition" between the two types (data set and image) due to the size of each kind of data. As a typical image is made of (440x300) 132000 pixels and has a similarity matrix between all the pixels of the order of (($13200^2$)/2) 8.5 billion elements, it is too big to be tackled in e.g. the R environment[16]. Thus some ways and tricks were/are to be found to

---

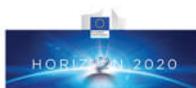

*This project is being funded by the European Union's Horizon 2020 research and innovation actions (RIA) programme under the grant agreement No 676036.*

adapt the clustering methods on images to extract their components.

In Section 4 we focused on the application of clustering techniques in astronomy. The astronomical community had long used the aforementioned traditional clustering techniques, primarily in the extragalactic domain at first to analyse the spatial distribution of galaxies and clusters of galaxies. Since the 1970's astronomers have developed a number of algorithmic clustering tools based on variations around these techniques and a review by Knebe et al. (2013) counted 38 algorithmic tools exist for halos detections. More recently new tools have been developed to detect filaments, voids and walls in the context of the cosmic web, but now also within the context of molecular gas analysis which is of particular interest in star forming regions.

In the context of the SFM project, we want to identify new measures and criteria from the observations to discriminate between the different star formation scenarios. On the one hand, our work aims to analyse the spatial/kinematics star distribution based on the identification of the most suited statistics or the derivation of new statistics to complement the clustering toolbox. On the other hand, we aim also at identifying the spatial/kinematics substructures. The study of the velocity dense subgroups will allow us to identify low versus high velocity stars and to study in a second step their location in their star forming region. The reverse of this study will also be worthwhile, i.e. to study the velocity distributions in spatially dense subgroups. If there are any dense subgroups in the phase space (i.e. the stars or young stellar objects located in a same spatial groups have also comparable velocities), would be also an interesting indicator that may, for example (and as a free speculation), suggest the local inheritance of the kinematics properties of the stars from the natal cloud.

A "good" clustering tool should be (1) robust to the free parameters and to the adding of new observations of the same feature; (2) reliable in retrieving same kind of features already identified and should fit into a theory or at least a model; and (3) to be part of a validation system that allow to make some predictions and interpretations of the current status. The first requirement of our work (analysing the spatial/kinematics of stellar distributions) implies a method that produces incomplete/partial clustering. The second requirement of our work (identifying the spatial/kinematics substructures) excludes all the K-means type methods since they derive only "roundish-ellipsoidish" convex structures.

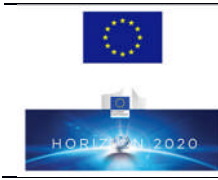

*This project is being funded by the European Union's Horizon 2020 research and innovation actions (RIA) programme under the grant agreement No 676036.*

These two requirements leave either the grid or density-based methods as our most viable options. Moreover, these two clustering methods both allow the handling of our third requirement, i.e. the hierarchy, even if we need to develop specific tools to do so. For the first requirement the use of the mutual nearest neighbour statistic and the one point correlation function is one of our propositions to outline the clustering properties at the smallest spatial scales (Joncour et al, 2017). For the second, we conclude that we need a method that can handle noise and outliers ("filtering"), arbitrary shapes and hierarchy.

Our ongoing investigation into these two trends has already produced promising results. Using the density based DBSCAN algorithm, with the free parameters set to such values appropriate to detect overdense region above the mean uniform random expectations at a 99.8% level of significance, reveal the smallest, densest and highly populated regions of Taurus, the archetype of the quiescent and the low-density star forming region (Joncour et al, in prep). In addition we have developed a promising tool to build a hierarchical "clustergram" that can handle a Nary-tree (the analogue of a dendrogram associated to a binary tree) based on a recursive DBSCAN. Our initial results are very encouraging, since the tool allows us to analyse the structure in terms of tree statistics (leaves, branches, etc.) which is a particular kind of the graph statistics.

To date we have mainly focused mainly on the spatial stellar clustering aspect. We intend to deal with the stellar kinematics in two ways: firstly we will study independently the kinematics clustering (radial velocity and/or proper motion) using the density-based algorithm to identify dense connected structures in the velocity space and then, secondly, we intend to study the spatial-kinematics structure, using the same clustering way in the phase space.

Our next study will be to compare the local molecular clouds properties (column density and radial velocity) to the ones of the young stars. One interesting idea is to compare their hierarchical structures, using the same clustering tools to derive clustergrams and dendrograms such that we can compare the tree statistics. We may also use correlation statistics to compare the stellar density map (obtained with the convolution by a kernel density function) and the column density of gas, to identify major trends and patterns.

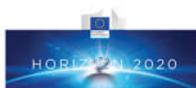 *This project is being funded by the European Union's Horizon 2020 research and innovation actions (RIA) programme under the grant agreement No 676036.*

To conclude, at this stage of the project we have already made our initial choices about the tools that we are currently investigating. It is too early to firmly establish whether we will stick with these choices or not, as this is dependent on the results we obtain. From our state of the art survey it is clear that there is no single clustering algorithm that will be able to handle all the issues, such as avoiding an exact requirement of domain knowledge, managing noise detection, arbitrary shape, large data sets, large dimensionality, and different types of data efficiently. Hence (perhaps unsurprising) there is no single method that would be optimal for all the analyses. The optimal method is the one that is capable to clearly answer the question that we submit to the data in order to constraint or reveal the nature of a physical phenomena. The clustering methods discussed in this report are a toolbox using smart techniques but they are not smart enough to automatically tell us which method to use. Performing clustering analyses requires a question to be answered to fully understand the underlying techniques and the structure of the data in order to adapt a strategy in choosing a particular tool. It requires an awareness of their strengths, flaws and their limits of reliability. It may also help to identify hidden patterns and in that sense these techniques help the discovery process. Our first results confirm these trends.

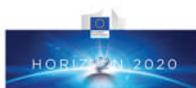

*This project is being funded by the European Union's Horizon 2020 research and innovation actions (RIA) programme under the grant agreement No 676036.*

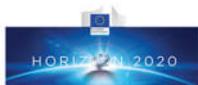

*This project is being funded by the European Union's Horizon 2020 research and innovation actions (RIA) programme under the grant agreement No 676036.*

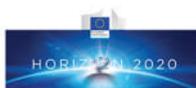
*This project is being funded by the European Union's Horizon 2020 research and innovation actions (RIA) programme under the grant agreement No 676036.*

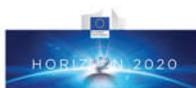

*This project is being funded by the European Union's Horizon 2020 research and innovation actions (RIA) programme under the grant agreement No 676036.*

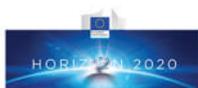

*This project is being funded by the European Union's Horizon 2020 research and innovation actions (RIA) programme under the grant agreement No 676036.*

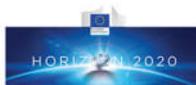

*This project is being funded by the European Union's Horizon 2020 research and innovation actions (RIA) programme under the grant agreement No 676036.*

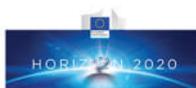

*This project is being funded by the European Union's Horizon 2020 research and innovation actions (RIA) programme under the grant agreement No 676036.*

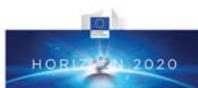
*This project is being funded by the European Union's Horizon 2020 research and innovation actions (RIA) programme under the grant agreement No 676036.*

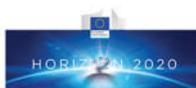
*This project is being funded by the European Union's Horizon 2020 research and innovation actions (RIA) programme under the grant agreement No 676036.*

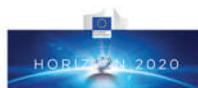

*This project is being funded by the European Union's Horizon 2020 research and innovation actions (RIA) programme under the grant agreement No 676036.*

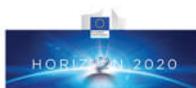

*This project is being funded by the European Union's Horizon 2020 research and innovation actions (RIA) programme under the grant agreement No 676036.*

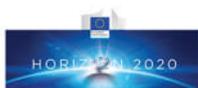

*This project is being funded by the European Union's Horizon 2020 research and innovation actions (RIA) programme under the grant agreement No 676036.*

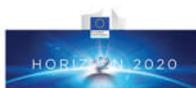

*This project is being funded by the European Union's Horizon 2020 research and innovation actions (RIA) programme under the grant agreement No 676036.*

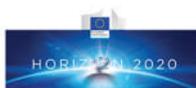
*This project is being funded by the European Union's Horizon 2020 research and innovation actions (RIA) programme under the grant agreement No 676036.*

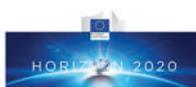

*This project is being funded by the European Union's Horizon 2020 research and innovation actions (RIA) programme under the grant agreement No 676036.*

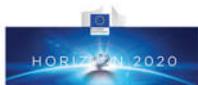

*This project is being funded by the European Union's Horizon 2020 research and innovation actions (RIA) programme under the grant agreement No 676036.*

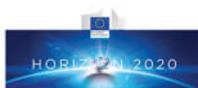

*This project is being funded by the European Union's Horizon 2020 research and innovation actions (RIA) programme under the grant agreement No 676036.*